\documentclass[range]{ar2e}%%%Put [range] option here for references citations to
\def\gtwid{\mathrel{\raise.3ex\hbox{$>$\kern-.75em\lower1ex\hbox{$\sim$}}}}
\def\ltwid{\mathrel{\raise.3ex\hbox{$<$\kern-.75em\lower1ex\hbox{$\sim$}}}}
\def\etal{et al.}
\def\spose#1{\hbox to 0pt{#1\hss}}
\def\simpropto{\mathrel{\spose{\lower 3pt\hbox{$\mathchar"218$}}
     \raise 2.0pt\hbox{$\propto$}}}

\begin{document}

%\input epsf.tex    %<-If you need EPS figures to be
                   %  called in {figure} environment for PC
%\input epsf.def   %<-If you need EPS figures to be
                   %  called in {figure} environment for Macintosh

\input psfig.sty

\jname{Annual Review of Astronomy and Astrophysics}
\jyear{2010}
%\jvol{}
%\ARinfo{1056-8700/97/0610-00}

\title{Thermal and Nonthermal Radio Galaxies}

\markboth{Antonucci}{Thermal and Nonthermal Radio Galaxies}

\author{Robert Antonucci
\affiliation{Department of Physics, University of California, Santa Barbara}}

%\begin{keywords}
%........
%\end{keywords}

\begin{abstract}
Radio galaxies usually show a subparsec-scale radio core sources, jets, and a pair of
giant radio lobes.  The optical 
spectra sometimes show only relatively weak lines of low-ionization ionic species,
and no clear nuclear continuum in the optical or UV region of the
spectrum.  Some show strong high-ionization narrow lines.  Finally, a few radio 
galaxies add broad bases onto the permitted lines.  These spectral catagories are the same as those for radio-quiet AGN and quasars.

By the 1980s, data from optical polarization and statistics of the radio
properties required that many narrow line radio galaxies produce do in fact produce 
strong optical/UV cotinuum.
hidden from the line of sight by dusty, roughly toroidal gas distributions.
{\it The radio galaxies with hidden quasars are referred to as ``thermal.''}
\par\vskip 2.5mm \indent Do all radio galaxies harbor hidden quasars? We
now know the answer using arguments based on radio, infrared, optical and X-ray
properties.  Near the top of the radio luminosity function, for FRII, GPS, and
CSS galaxies, the answer is yes. Below the top of the radio luminosity
function, many do not. At low radio luminosities, most do not. Instead these
{\it ``nonthermal'' weakly-accreting galaxies} manifest their energetic output
only as kinetic
energy in the form of synchrotron jets.  These kinetic-energy-only ("nonthermal")
radio galaxies are a subset of those with only weak
low-ionization line emisison.  This applies to all types of radio galaxy, big FR II
doubles, as well as the small young GigaHertz-Peaked-Spectrum and Compact Steep
Spectrum sources. Only a few FR I sources are of the thermal type.
\end{abstract}

\newpage

\maketitle

\section{TERMINOLOGY}

\subsection{Optical/UV Classes of Radio Sources}

This paper concerns primarily radio loud active galactic nuclei (AGN). 
``Radio loud''
is sometimes defined by an absolute radio luminosity cutoff, and 
sometimes (less
usefully) by a ratio of radio to optical luminosity. The nomenclature is
mutifaceted, complex, and very confusing for a newcomer. We will divide 
the AGN
into two broad classes, which correspond to the two popular and persuasive
central-engine models.  The presence of an optical/UV
continuum of the type called the Big Blue Bump will be called ``thermal''
because there is a consensus that this is thermal radiation from a copious
opaque and probably usually geometrically thin accretion flow.  This 
includes the radio
loud quasars, the broad Broad Line Radio Galaxies, and the objects that
have similar accretion flows hidden from the line of sight.
(Some papers define the Big Blue Bump as the excess over a
notional power law extending from the near-IR to the far-UV or X-ray, 
but that
is not the most common usage, or the present usage.) The Big Blue Bump is
virtually always accompanied by conspicuous broad permitted emission
lines\footnote{Possible exception to this one discussed in the next 
section.} from
regions collectively called the Broad Line Region. This combination is 
called a
Type 1 spectrum.

By contrast, a radio loud AGN which lacks visible broad lines,
is called a Narrow Line Radio Galaxy (``Type 2'' optical spectrum). The
narrow-line spectra of all radio types vary enormously from optically 
weak Low Ionization
Galaxies --- sometimes loosely called Low Excitation
Galaxies --- like M87, to very powerful High Ionization Galaxies like 
Cygnus A.
The former are turning out to be almost all ``nonthermal'' radio 
galaxies, lacking a
powerful Big Blue Bump and Broad Line Region, even a hidden 
one.\footnote{Note
that there is only a little evidence yet that the thermal and nonthermal
objects are bimodal in any property, and such isn't necessarily expected
theoretically.}

If the Big Blue Bump is directly visible in the total-flux spectrum, the 
object
is called a radio loud quasar,\footnote{Recall that this paper is largely
restricted to radio loud AGN.} or if low in optical luminosity
(e.g. $M(V) > -23$ for $H_0= 50$ km sec$^{-1}$ Mpc$^{-1}$, as adopted 
for the older Veron-Cetty
and Veron catalogs), it may be called a Broad Line Radio 
Galaxy.\footnote{It is
sometimes argued that at these low luminosities, there are some relatively
subtle differences with respect to the quasars, e.g., van Bemmel and Barthel
2001, but such a distinction will not be made here.}\footnote{For 
redshifts above
a few tenths, the first two words of ``Narrow Line Radio Galaxy'' are 
often dropped,
because the broad line objects are unambiguously called quasars.} In 
fact when
it's clear from context, ``quasars'' will be taken to include Broad Line 
Radio Galaxies.

For radio-bright objects at redshifts of larger than a few tenths, the 
presence
of a (directly visible or hidden) optical/UV Big Blue Bump
is general --- except in those rare objects whose optical/UV spectrum is
overwhelmed by beamed synchrotron emission from the bases of favorably
oriented relativistic jets (``Blazars'').
For objects with a large contribution to the optical/UV continuum by
highly variable, highly polarized beamed synchrotron radiation, the
general term is Blazars, defined in Stein 1978.
Blazars are defined as the union of two
classes: 1) objects in which a Big Blue Bump/Broad Line Region is still 
discernable against
a strong synchrotron component (Optically Violently Variable Quasars, also
known as Highly Polarized Quasars) with 1960s-1970s technology, and 2)
objects with a pure synchrotron continuum in those old spectra, and
little or no detectable line emission or absorption (BL Lac Objects).
However, it's been known since the 1970s at least that many
historically defined ``BL Lacs'' show emission lines, both narrow and
broad, especially (but not necessarily) when observed in low states. For
example, BL Lac itself has weak narrow emission lines, and stellar
absorption lines (Miller 1981); now we know
that broad lines are often visible as well (Vermeulen et al 1995). In fact
it is well known that many highly polarized, violently variable quasars
are indistinguishable from BL Lacs when in high states (Miller and French
1978; see also Miller 1981).

None of these, ``BL Lac'' nor ``High Polarization Quasar,'' nor
``Optically Violently Variable Quasar'' is very well defined, and many
studies have been damaged by blindly using these historical categories
--- sometimes just from catalog classifications, or by trying to mimic
them with equivalent width cutoffs, which result in classifications
changing with time (Antonucci et al 1987, 2002a)! It is
also still often incorrectly asserted that the parent population
(equivalent objects at more than a few degrees inclination) for ``BL
Lacs'' is FR I double radio sources\footnote{FR I and II radio sources are
discussed in Sec.~1.3.} (the lower-luminosity edge-darkened
ones). This is manifestly not the case (see references to maps of
diffuse radio emission in Antonucci 2002a, going back for decades).
Yet people still write about FR II radio emission in BL Lacs as a
``problem'' for the unified model. (It is well known that parents of
optically-defined ``BL Lacs'' can be of either FR type, e.g. Kollgaard et al
1992.) This has invalidated studies of cosmological evolution, among other
things (e.g., Ostriker and Vietri 1985, 1990.) See Jackson and Wall 1999 
for a
well-informed and sensible discussion.

Great care is required in classifying AGN, and the price of carelessness is
spurious results. For example, historically 3CR382, 3CR390.3 and similar 
objects
were called Broad Line Radio Galaxies (Type 1 optical spectrum), and this is
still reasonable. But the same was done for 3CR234, because in fact the
broad H-$\alpha$ line is visible in the total-flux spectrum. This wasn't
an error in its historical context, but we now know that the broad
lines and Big Blue Bump are seen only by reflection (Antonucci 1982, 
1984; Tran
et al 1995; Young et al 1998). Therefore from the point of view of
unified schemes, such objects must be included with the Narrow Line
Radio Galaxies (Type 2 optical spectrum), just as we keep Seyfert 2s (with
hidden Big Blue Bump/Broad Line Region) separate from the Seyfert 1s. It 
would
be great if a multi-dimensional classification scheme could be shown in a
drawing, but a confusing ``tesseract'' would be needed (Blandford 1993). 
Even
a tesseract might oversimplify the true situation.

\subsection{AGN with Big Blue Bumps, but said to lack Broad Emission Lines}

Note:  this section is rather technical, and it may be skipped by the general
reader.

Laor and Davis' (2011) modeling paper has a convenient list of
observational papers reporting on very, very few quasars with Big Blue Bumps,
yet little or no detectable broad line emission.  One limitation of these
analyses is that any equivalent width upper limits rely on making an assumption
about the broad line width.  This is what led Stockton et al (1994) astray in
claiming that the radio galaxy Cygus A has no broad emission lines.  See
Tadhunter et al 1990; Antonucci et al 1994, and Ogle et al 1997 for correction
of the Stockton et al claim.  However, extremely large line widths are
exceptional for quasars, and do not occur in the faint but detectable broad
lines in some of these very weak-lined quasars, so the limits on broad line
equivalent widths are probably fairly close to the truth.

Laor and Davis (2011) also cite a few
papers on a very different kind of AGN.  These papers claim (dubiously in my opinion:
Antonucci 2002b) that for a few nearby AGN, there is no detectable broad line
emission, at an interesting level.  The authors refer to these as ``True Type 2
Seyferts," because broad lines are undetectable in both total flux and polarized
flux.

Some problems with these claims:  1) in total flux, the upper limits on broad
line equivalent widths depend on assumed line widths in most cases;
2) there are
no upper limits given for broad line equivalent widths in polarized flux.
Note that such limits must be calculated {\it relative to the BBB component
of the continuum only}, and this is hardly possible because the spectra in these
objects are very much dominated by starlight from the host galaxies;  also there
is no prediction available for the flux of scattered light from a putative
hidden
nucleus, detection of which requires well-placed scattering ``mirrors."
Also the polarized flux from a large fraction of Type 2 AGN comes from
transmission of the optical light starlight in the host galaxy in many cases
(Thompson and Martin 1988), so even if upper limits were given on the line
equivalent widths in polarized flux for these objects, they would be irrelevant.

\section{BROADEST SKETCH}

\subsection{Where To Find Basic Information}

Active Galactic Nuclei (AGN) is a term encompassing a variety of energetic
phenomena in galactic centers which are thought to be powered directly or
indirectly by accretion of matter onto central supermassive black holes. 
There
are many reviews of this field, including several of book length. 
Perhaps the
most technical and broad book-length review is that of J.~Krolik, which was
published in 1998. The classic text on spectral analysis of gaseous nebulae
and AGN has been updated in 2006 (Osterbrock and Ferland); it's a wonderful
book, but it is biased towards the optical and ultraviolet regions of the
spectrum. A new edition (2009) of An Introduction to Radio Astronomy by 
Burke
and Graham-Smith is also noteworthy.

Few theoretical predictions have been borne out in this field, and 
understanding
is still semi-quantitative at best (Alloin et al. 1985; Antonucci 1988, 
2002a;
Courvoisier and Clavel 1991; Koratkar and Blaes 1999; Blaes 2007). This 
review
will refer to some fairly general theoretical ideas, but it will mainly 
organize
some observational information on radio galaxy central engines that has 
become
clear over recent years. It will not include a general introduction to 
AGN (see
above references), but will address the nature of the central engines in 
radio
galaxies which is a topic tied up as a practical matter with orientation 
effects
on observations.  Orientation effects are introduced in the next 
section, and as
needed throughout the text.

The general topic of orientation effects (``Unified Models'') is reviewed in
detail in Antonucci 1993. The material in that review is almost entirely
``still true.''  However, it has been updated and elucidated in
several more recent (but generally narrower) reviews (Urry \& Padovani 1995;
Dopita 1997; Cohen et al. 1999; Wills 1999; Axon 2001; Tadhunter 2008).

In a nutshell, prior to the mid-1980s, it seemed that radio loud quasars 
(with
their powerful thermal optical/UV light) were quite distinct from radio
galaxies; the latter are analogous to quasars in general radio 
properties, but
apparently lacked the strong optical/UV electromagnetic luminosity (Big Blue
Bump). In a landmark review of bright extragalactic radio sources, Begelman,
Blandford and Rees (1984) posited that {\it ``The ratio of mass 
accretion rate
to the mass of the hole may determine whether a [radio loud] active galactic
nucleus will be primarily a thermal emitter like an optical quasar or a
nonthermal object like a radio galaxy.''}

Through optical spectropolarimetry and other means, it was subsequently
determined that many radio galaxies, especially the most powerful ones,
with strong high-ionization narrow emission lines, actually harbor hidden
quasars surrounded by opaque dusty tori, so that the
observational appearance depends on the inclination of the radio axis to the
line of sight. But now we know that for many radio galaxies, any hidden 
quasar
must be very weak.

The radio quiet and radio loud objects with high ionization are 
virtually all
visible (Type 1) or hidden (Type 2) Seyferts or quasars (e.g., Antonucci 
2002b).
At lower radio luminosities of all radio types, we find mostly 
LINERs\footnote{A
few LINERs have strong broad lines, e.g., Filippenko and Halpern 1984. 
Many LINERs
have very inconspicuous broad H$\alpha$ components, but it isn't clear 
to me that
they are strictly analogous to those in low-luminosity Seyferts. For 
example, we
do not know whether they vary rapidly (L.~Ho, 2011, private communication).}
(Low Ionization Nuclear Emission Regions). A recent
and comprehensive review of LINERs is that of Ho (2008).

\subsection{Nature of Geometrical Unified Models}

In the low-redshift universe, there is a near-perfect correspondence between
``radio loud'' objects --- with $L_\nu$(1.5 GHz) loosely extending from 
perhaps
$\sim10^{28}$--$10^{36}$ erg sec$^{-1}$ Hz$^{-1}$ --- and elliptical 
hosts. (This
paper uses H$_0=70$ km sec$^{-1}$ Mpc$^{-1}$, $\Omega_{\rm matter} = 
0.3$, and
$\Omega_\Lambda = 0.7$.) For a fiducial $L_\nu\simpropto\sim\nu^{-1}$ 
spectrum,
for which there is equal power per logarithmic frequency interval --- also
referred to as ``per dex'' --- the parameter $\nu L_\nu$ gives the power
integrated over an interval of 0.30 dex (powers of 10). Thus if one 
integrates
such a spectrum over the ``radio region'' 30 MHz--300 GHz, the corresponding
luminosity is 13.3 times as great; a look at various radio AGN in the NASA
Extragalactic Database (http://nedwww.ipac.caltech.edu) shows that the
luminosity per dex of the dominant optically thin synchrotron component 
tends
to be lower outside this range, so depending on the application, one may
consider this as a crude ``radio bolometric correction.'' Resulting 
radio powers
can exceed $1\times10^{45}$erg/s. Estimates of energy tied up in 100-kpc 
scale
radio lobes are as high as $1\times10^{61}$ ergs or more, even using the
particle/magnetic field minimum-energy assumption and assuming a lack of a
dominant proton contribution.  (The minimum-energy assumption posits 
that energy
is apportioned between relativistic electrons and magnetic field in such 
a way
as to minimize lobe energy content for a given synchrotron luminosity.)

\begin{minipage}{5in}
\vskip 4.5truein
\includegraphics{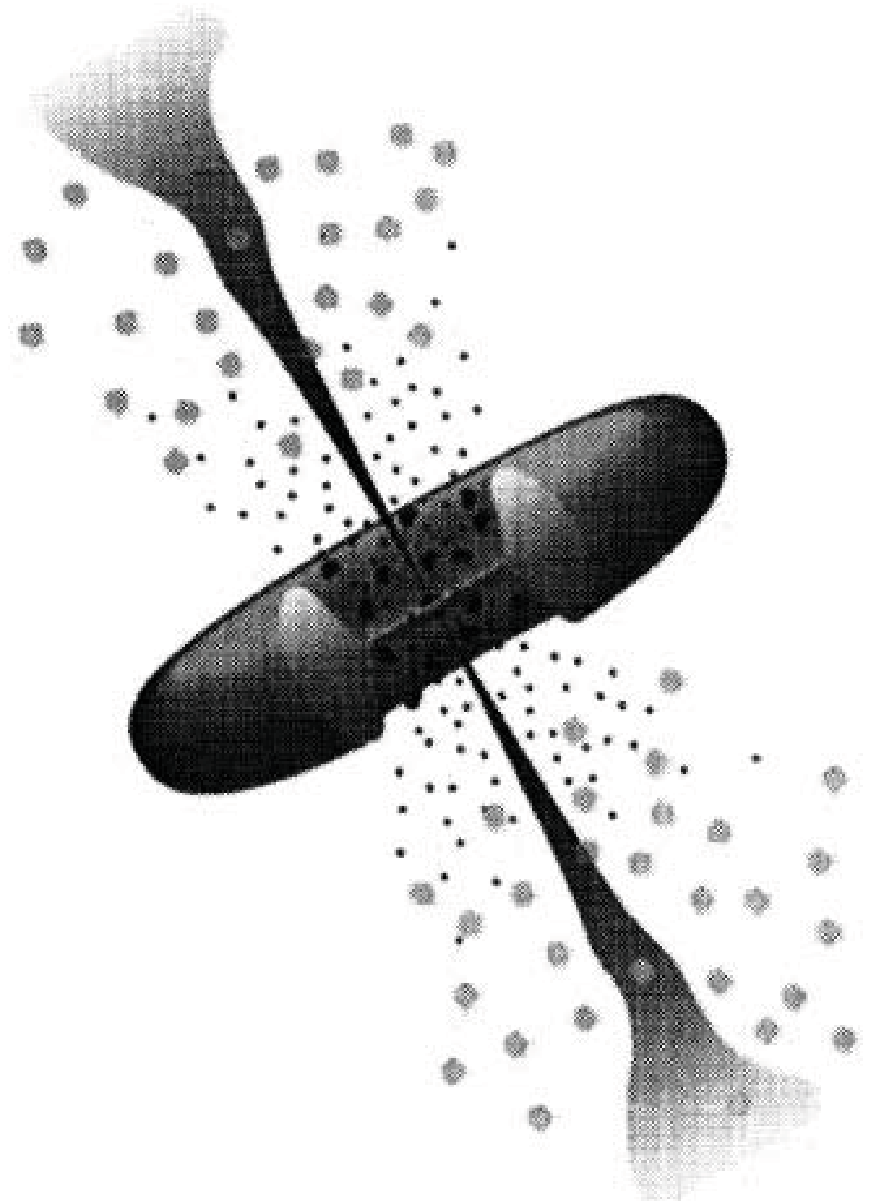}
\setlength{\baselineskip}{1ex}
\vskip 4mm
{Figure 1. A schematic diagram of the current paradigm for radio-loud 
AGN (not
to scale). Surrounding the central black hole is a luminous accretion disk.
Broad emission lines are produced in clouds orbiting outside the disk and
perhaps by the disk itself. A thick dusty torus (or warped disk) 
obscures the
Broad-Line Region from transverse lines of sight; some continuum and 
broad-line
emission can be scattered into those lines of sight by warm electrons or 
dust
that are outside the observing torus. Narrow emission lines are produced in
clouds much farther from the central source. Radio jets, shown here as the
diffuse 2-sided jets characteristic of low-luminosity, or FR I-type, radio
sources, emanate from the region near the black hole, initially at 
relativistic
speeds. (adapted from Urry \& Padovani 1995)}
\end{minipage}
\vskip 2.5mm

Unified models assert that certain AGN classes differ only in 
orientation with
respect to the line of sight. These models comprise two separate (though
interacting) assertions, as illustrated in Fig.~1. The first to be 
recognized
historically is the effect of relativistic beaming (aberration causing
anisotropy in the observed frame) in the powerful synchrotron jets which 
feed
particles and magnetic energy into the radio lobes. When seen at low
inclinations, beaming amplifies and speeds up ``core'' (subparsec scale jet,
usual synchrotron-self-absorbed)
radio flux variability and (apparent faster-than-light) ``superluminal
motion.'' Thus a special fortuitous orientation of a nearly axisymmetric 
object
leads to a different observational category (Blazars). The same objects, 
seen
at higher inclination, are ordinary radio-loud galaxies and quasars 
(Blandford
et al 1984; Antonucci and Ulvestad 1985; Kollgaard et al 1992, etc.). We can
call this the beaming unified model (or more loosely, just the beam model,
though that term usually connotes some connection with actual physics).

At low redshift, radio-quiet (but not silent) AGN lie in spiral hosts, 
and go
by the name of Seyfert galaxies. They rarely show detectable motions in 
their
weak radio jets --- and when they do show motions, the apparent speed is 
usually
much less than the speed of light. More luminous radio quiet objects are 
called
Radio Quiet Quasars, or historically, Quasistellar Objects (QSOs).

But both radio quiet and many radio loud AGN widely exhibit another kind of
orientation unification: many well-studied objects include energetically
dominant continuum components in the optical-ultraviolet region, 
referred to as
the Big Blue Bump, and widely attributed to thermal radiation from 
optical thick
accretion flows. (Confirmation of the latter can be found in
Kishimoto et al 2004, 2005, 2008.) They are almost always accompanied by 
broad
(5000--10,000 km/s) permitted emission lines. Both these components 
reside inside
optically opaque dusty structures which to zeroth order have the shadowing
properties of tori, and these structures are referred to loosely as the 
``AGN
torus.'' In some cases we have a direct (low-inclination, polar) view of 
these
compact components (in quasars, Seyfert 1 galaxies, and Broad Line
Radio Galaxies). In many other cases (high inclination, equatorial 
views), we
can be sure these two components are present inside the tori using the
technique of optical spectropolarimetry, which literally allows us to 
see the
nuclei ``from above,'' using ambient gas and dust as natural periscopic 
mirrors.

This wonderful trick of optical spectrapolarimetry (e.g. Antonucci 1982, 
1983,
1984; Antonucci and Miller 1985) uses the polarization property of scattered
light to separate the spectrum of hidden sources from any sources of direct
light. The above references show that the Big Blue Bump and the Broad Line
Region are present but hidden from direct view. The scattering polarization
position angle indicates that the photons emerging from these components can
only escape from the hidden nuclei if moving along the radio jet (and lobe)
axis, so that the other directions must be blocked by an obscuring 
torus. This
is illustrated in Fig.~2, from Tran 1995, illustrating the total and
polarized fluxes, for 3CR234. The polarized photons (scattered light) 
contribute to
both plots of course, but show up with great contrast when plotted alone.

\begin{minipage}{5in}
\vskip 2.5truein
\includegraphics{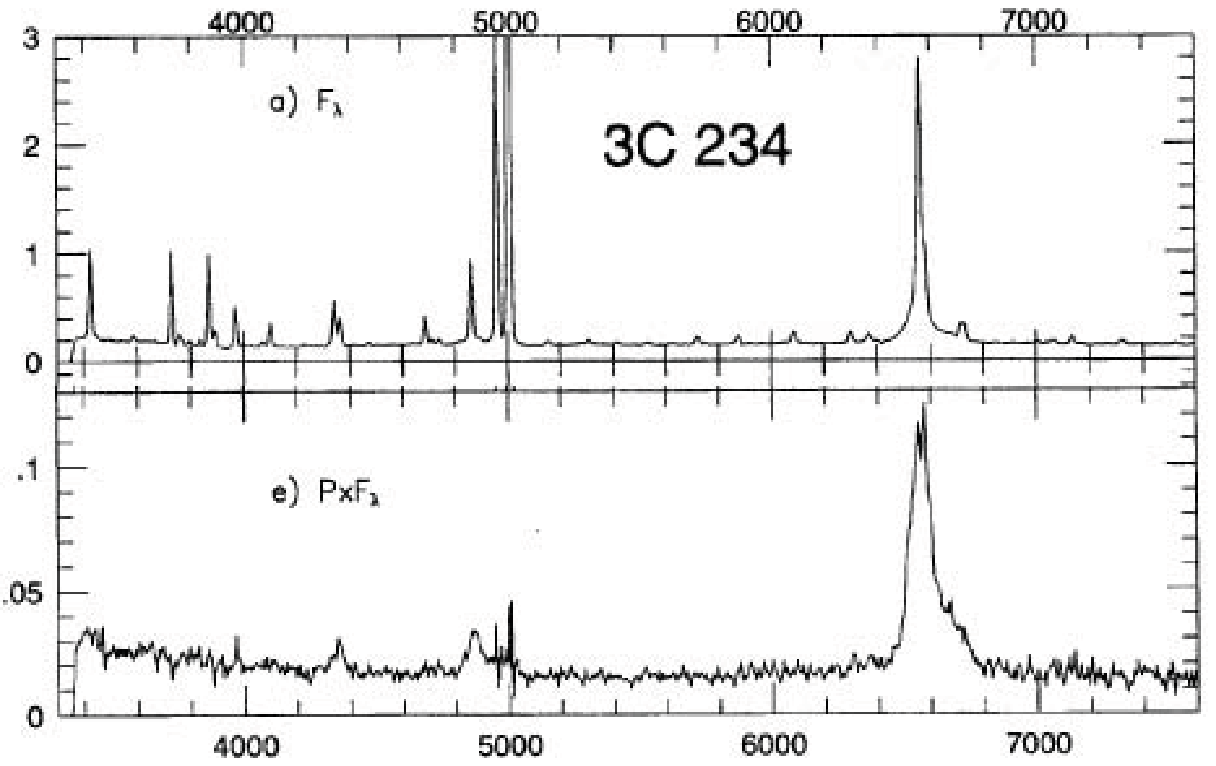}
\setlength{\baselineskip}{1ex}
\vskip 4mm
{Figure~2. Total and polarized flux for the Narrow Line Radio Galaxy 
that showed
as the first hidden quasar, 3CR234 (Antonucci 1984; the figure shows the 
data of
Tran \etal\ 1995.). The polarization angle relative to the radio axis 
shows that
photons can only stream out of the nucleus in the polar directions. The 
polarized
flux (akin to scattered light) spectrum shows the hidden quasar features 
at good
contrast. (In this case, however, the scattered ray is itself reddened.)}
\end{minipage}
\vskip 2.5mm

If all objects fit in with the above descriptions of orientation 
unification, we
would be left with as few as two independent types of ``central engines'' on
relativistic scales\footnote{This refers to regions where the jet 
velocity is
almost the speed of light, $c$, and also the gravitational potential 
energy per
unit mass is of order $c^2$.}, those for radio
quiet and those for radio loud AGN.  But we now know that many radio 
galaxies
lack powerful visible or hidden Type 1 engines (Big Blue Bump, Broad Line
Region, copious accretion flows). They occur frequently in some regions of
parameter space (as delineated for example by redshift and radio flux or
luminosity), and not in others. The purpose of this review is to gather the
multiwavelength evidence for the last two statements, and emphasize the
dependence of the distribution of central engine types on parameter space,
which is the key to avoiding many errors and much confusion and even 
discord.
We will see that a great deal of self-consistent information is available on
the occurrence of the two types of radio galaxy/quasar central engines.

\subsection{Types of Powerful Radio Source; Scope for Unification by 
Orientation}

Within the radio loud AGN arena, several types are also distinguished 
based not
on central engine properties but on extended radio morphology. 
Unfortunately,
they do not correspond very closely to the optical catagories!

The most luminous giant ($\ell\gtwid100$ kpc) double radio sources are
designated FR II (``Classical Double''), for Fanaroff and Riley (1974); 
those
authors noted that a whole suite of properties change together fairly 
suddenly
over a critical radio luminosity (referring to the roughly isotropic diffuse
emission). Sources above $\sim2\times10^{32}$ erg/sec/Hz at 1.5GHz show
edge-brightening and hot spots, where the radio jets impinge on an external
medium, and shocks partially convert bulk kinetic energy to particle and 
field
energy. They also tend to have strong side-to-side asymmetry (generally
attributed to relativistic beaming) of the jets over scales from the 
relativistic
region up to tens of kpc. The lower luminosity giant (also 
$\ell\gtwid100$ kpc)
objects (FR I galaxies) also have strong side-to-side jet asymmetry, but 
only on
1--1000pc scales in most cases. An important refinement to the FR 
classification
scheme is that the dependence of the exact radio luminosity cutoff 
depends on
the optical luminosity of the host galaxy (Owen and Ledlow 1994, but see 
Best 2009, Fig.~4a).

The ``FR'' types of radio galaxy generally have sizes of 25--1000kpc, 
but large
populations of smaller sources exist, and they can still be very 
powerful. (The
small sizes mean that their lobe energy content is very much lower, 
however.)
They are denoted in an inconsistent way, according to their means of 
discovery.
Optically thin, steep-spectrum radio sources which were historically 
unresolved
on arcminute scales, were (and are) called Compact Steep Spectrum
(CSS) sources to distinguish them from opaque beamed synchrotron cores; 
their
spectra peak at $\sim100$MHz, and they are generally defined to be in 
the size range
1kpc--15 or 25kpc. Sometimes they are crudely defined to be sources smaller
than typical host galaxies. Sources whose extent is less than 
$\sim1$kpc, with even more compact substructure, are
often dominated by synchrotron components which are self-absorbed up to 
$\sim$GHz
frequencies\footnote{It is proposed by Begelman (1999) that the weak 
fluxes at
low frequencies result from free-free rather than inverse synchrotron 
absorption.
Encouraging follow-up work can be found in Stawarz et al 2008 and 
Ostorero et al
2010.}, and are called Gigahertz-Peaked-Spectrum sources (GPS). (Even
tinier sources are being sought by selecting for self-absorption peaks 
at even
higher frequencies.) These classes are reviewed by O'Dea (1998). Since
that paper was written, much evidence has accumulated from VLBI proper 
motions
that the sources are small because they are very young ($\sim1000$--100,000
years!\footnote{This material was reviewed recently by Giroletti 2008.}).
In at least some cases it is known from faint extended emission that 
these very
young ages refer only to a recent phase of activity however. Statistically,
only a very small fraction of small and very short-lived sources can 
grow to be
huge bright long-lived sources.

The relevant properties of classes will be discussed in turn, generally 
starting
with the radio data and proceeding upwards in frequency. We will 
discover that
some FR II radio galaxies at $z\ltwid0.5$--1.0 lack powerful hidden quasars.
These objects {\it may} have hidden Type 1 nuclei, but they are constrained
to be much weaker than those of the ``matched''\footnote{In radio flux 
and redshift.} visible quasars, and thus they
are not ``unified'' (identified) with them via orientation with respect 
to the
line of sight. At low redshifts ($z<\sim0.5$) it is probable that only a 
minority
of FR II radio galaxies in the 3CR catalog host hidden uasars.

Next we will tackle the less powerful (FR I) giant radio galaxies, which 
are by
selection nearby in almost all cases. The 3CR catalog flux cutoff of 10 
Jy at
178MHz (Laing, Riley and Longair 1983) corresponds to the FR I vs.\ II radio
luminosity separation at $z \sim0.2$. Most (but by no means all) of these
objects have nuclear spectral energy distributions dominated by synchrotron
radiation, with no evidence for visible or hidden ``Type 1'' central 
engines.

Finally the small, young GPS and CSS sources will be discussed. Very recent
information from the ISO and Spitzer infrared satellites has greatly 
increased
our knowledge of ``shadowing unification''\footnote{As a reminder, 
unification of broad line and
narrow line AGN by orientation of a toroidal nuclear obscurer is here called
``shadowing unification.'' Ascription of superluminal motion and other
relativisitic effects in subpopulations to orientation is called ``beaming
unification.''} at various radio luminosities.

A major caveat of this paper, and of this field, is that most of the
information derives from the brightest radio sources, especially those 
in the
3CR catalog, so no implication is made for unexplored regions of parameter
space! Another major caveat is that while we discuss thermal vs.\ nonthermal
galaxies\footnote{Recall that thermal vs.\ non-thermal refers not to the 
radio
emission itself, but to the presence of an energetically dominant 
optical/UV source thought to
arise from an accretion disk.}, relatively little evidence is presented that
any parameter is bimodal, so that there could be a continuum of properties.

\subsection{The Infrared Calorimeter}

Radiation absorbed by the dusty torus is largely reradiated as infrared, and
many studies have concluded that in reasonably luminous AGN (so that the 
IR is
not dominated by a normal host galaxy), at least the near- and mid-infrared
reradiation ($\approx1$--40 microns) is dominated by reprocessed nuclear
optical/UV/X-ray light. In specific populations, the entire IR seems to be
radiated by the torus, because the colors are warm throughout, and there is
evidence for only weak star formation (e.g. PAHs) or synchrotron radiation
(e.g., strong radio-mm emission). Thus the infrared reradiation of 
nuclear light
can potentially be used as a {\it calorimetric indicator} for the
luminosity of any hidden AGN.

There are at least two ambiguities in using
this infrared emission as an AGN calorimeter. The first is that some of the
nuclear radiation may not reach the torus, either because of the intrinsic
latitude-dependence expected for many models of the Big Blue Bump
(e.g. Netzer 1985), or preferentially planar Broad Line Region absorption
(Maiolino et al 2001c; Gaskell et al 2007). Both of these effects add 
noise to
the dust reradiation calorimeter, and tend to make hidden quasars look 
dimmer
than visible quasars in the infrared for a given opening angle.
Nevertheless most studies show infrared luminosity about as expected, from
detailed studies of individual objects (Carleton et al 1984; 
Storchi-Bergmann
et al 1992) and for populations of Type 1 and Type 2 objects within
isotropically selected samples (Keel et al 1994). Carleton et al (1984, 
see Fig.~3) shows how the
infrared calorimeter works, based on the first spectropolarimetric 
hidden AGN, 3CR234.

\begin{minipage}{5in}
\vskip 2.5truein
\includegraphics{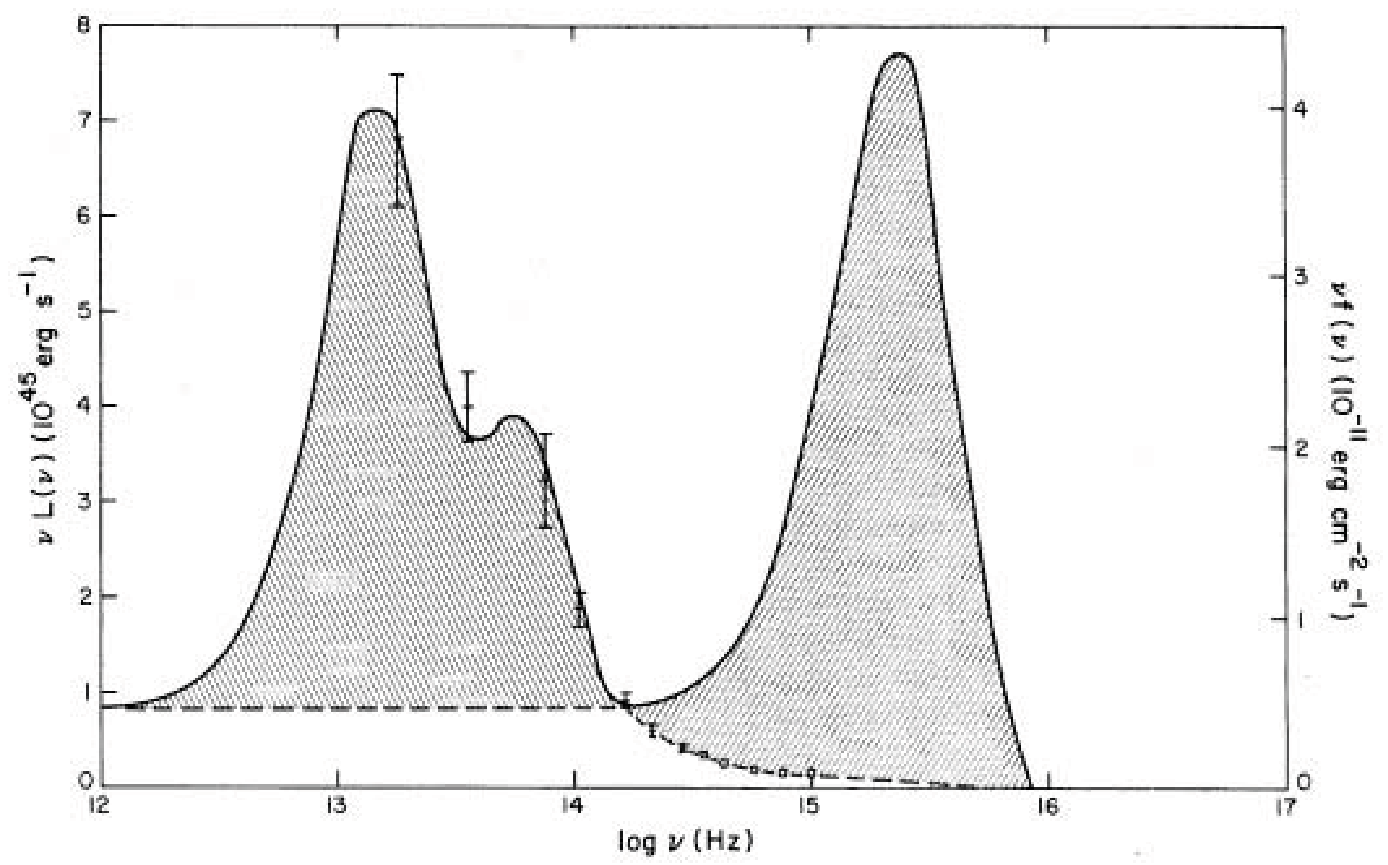}
\setlength{\baselineskip}{1ex}
\vskip 4mm
{Figure~3. 3CR234 continuum observations plus model fits. The ordinate 
is linear
in $\nu L_\nu$, and the area under any portion of a curve is 
proportional to the
luminosity in that portion of the spectrum. The hidden
optical/UV component (Big Blue Bump) has been reprocessed into the 
infrared in
this hidden quasar.}
\end{minipage}
\vskip 2.5mm

Also, although the covering factor deduced from dividing infrared 
re-emission
by Big Blue Bump luminosities are therefore lower limits, they tend to 
be high
($\sim0.1$--1) so that they can't be too far off!\footnote{Although Maiolino
et al 2007 didn't actually integrate over the SED and should therefore be
viewed with much caution, it is interesting that they actually find some
covering factors nominally greater than one for low-moderate luminosity 
objects.
This was predicted qualitatively by Gaskell et al (2004), who argued that a
preponderance of large grains leads to less apparent absorption (spectral
curvature) in the optical/UV region than otherwise, and thus one may 
fall into
the trap of inferring very low extinction. The inner torus is expected 
to lack
small grains, most robustly perhaps on the grounds that the torus 
sublimation
radius must be larger for small grains according to radiative 
equilibrium. The
lack of small grains at the sublimation radius is supported by the data of
Suganuma et al 2006: see discussion in Kishimoto et al 2007.}

Another concern with the infrared calorimeter is the expected anisotropy 
of the
thermal dust emission due to the large dust column densities (Pier and 
Krolik
1992, 1993).  Many Type 2 AGN have X-ray columns of $\gtwid 1 \times 
10^{24}$
cm$^{-2}$; absorption of mid-IR lines, molecular maps, and the great
difference in the average X-ray columns between Type 1 and Type 2 AGN 
suggest
that a commensurate dust extinction is present. (See Maiolino et al 
2001a,b,c
for arguments which affect this line of reasoning quantitatively but not 
qualitatively.)

There is a limit on the anisotropy of the ratio [O III] 
$\lambda$5007/F$_\nu(60\mu)$
in Seyferts from Fig.~3 of Keel et al 1994. The figure shows that in their
well-selected sample ($60\mu$ flux with a mild $25\mu$--$60\mu$ warmth 
criterion),
Type 1 and Type 2 objects have indistinguishable distributions of 
L($60\mu$),
L[5007], and of course their ratio. The $\lambda$5007 line is produced 
outside
the torus in most objects like these: it doesn't appear in polarized 
flux along
with the broad emission lines and Big Blue Bump. Thus it's not significantly
hidden inside the torus, is likely quite isotropic in this parameter 
space, and
is nearly isotropically selected\footnote{Isotropic selection means 
selection on
an isotropic AGN property, which avoids powerful biases in comparisons 
between
classes. It is essential in order to produce intelligible results.}, 
fairly powerful Seyferts.

Torus models do predict that the optical depths will be small in the 
far-infrared
in general. For AGN-dominated infrared SEDs, that means there is an elegant
and detailed method of deriving the degree and wavelength-dependence
of the dust emission anisotropy. For isotropically selected samples, we can
divide composite or representative Type 1 SEDs by those for Type 2, 
tying them
together at $60\mu$.

In order to make the infrared calorimeter accurate, we need to account 
for the
common anisotropy of the near- and mid-thermal dust emission. There is 
general
agreement on near isotropy past $\sim30\mu$, as was first predicted by 
Pier and
Krolik (1992, 1993). The main basis for their prediction of anisotropy at
shorter wavelengths was that many Seyfert
2's are Compton-thick, including NGC1068 --- which is completely opaque 
and for
which $N(H)\gtwid10^{25}$ according to the X-ray spectrum (e.g., Pounds 
and Vaughan
2006). In Galactic dusty gas this corresponds to $A(V) \sim 1000$. (See 
e.g.,
Maiolino et al 2001a,b for a quantitative correction, which however 
doesn't greatly
affect the discussion of gas vs.\ dust columns below.) As an aside, I do 
think
that dust-free atomic gas can contribute to the X-ray absorption in some 
cases
(e.g., Risaliti et al 2011; Antonucci et al 2004), but the X-ray columns of
Type 2 AGN are on average $\gtwid100\times$ those of Type 1, so most of 
the column is
connected with the Type 2 classification. Also in some cases, the dust 
column
can be constrained to be similar to the very high X-ray columns. For 
example,
Lutz et al (2000) used the lack of Pf-$\alpha$ at $7.46\mu$ to show that
$A(V) > 50$ in NGC 1068.

Recall that one could pick spectral Type 1 (broad-line) radio loud 
quasars and
radio galaxies by some fairly isotropic AGN-related luminosity (e.g.,
lobe-power), and divide the two spectra to produce a spectrum of diminuation
from anisotropy. We were able to do this for the $z>1$ 3CR (Fig.~4) 
because all
of the radio galaxies seem to host hidden quasars, but only out to 
$\sim15\mu$
in the rest frame. {\it The correction for anisotropy is around a factor
of 10 in the near-IR, 1.5--2 on both sides of the silicate feature,
and about three in the silicate feature}, but varies significantly from 
object
to object. As expected, it flattens out at $\sim1$ at long wavelengths.

Recall that torus shape was inferred from the high\footnote{A common
error is to use the percent polarization of the continuum (after
correcting for bulge or elliptical starlight) as the scattering 
polarization:
that is usually contaminated by hot stars, and only the
broad lines themselves can be used to find, or more often place a
\underbar{lower limit} on, the percent polarization of the scattered
nuclear light (Antonucci 2002b).} typical polarization of the
reflected light in hidden-Broad Line Region objects, which is generally 
perpendicular
to the radio axis, meaning that photons can only escape the nucleus to 
scatter
into the line of sight if they leave the nucleus in the polar directions.

The obscuring dusty tori invoked in the shadowing aspect of the unified 
model
are active, not just passive, components. Modulo factors of order unity for
geometry and dust cloud albedo, the tori will reprocess almost all of the
incident Big Blue Bump/Broad Line Region luminosity into the infrared. 
Thus the ratio of re-emitted
infrared emission to Big Blue Bump (and often considerable absorbed 
X-ray) emission tells
us the approximate covering factor of the dusty gas which is idealized as a
torus shape.\footnote{There is at
least one obscured quasar with a \underbar{thin} obscuring disk --- the
radio-loud, mini-MgII BAL OI287 (Goodrich and Miller 1988; Rudy and Schmidt
1988; Ulvestad and Antonucci 1988; Antonucci, Kinney, and Hurt 1993); 
however, even
that one seems to have an upturn longward of $1\mu$ at least according 
to the
NED figures.}

The covering fraction should agree with the fraction of Type 2 (hidden) 
nuclei
in a sample which is selected by any isotropic AGN property. What are the
results of this comparison? I'm crudely leaving out intermediate types, 
and the
substantial minority of Seyfert 2s whose Type 1 nuclei are hidden by 
dust in the
host galaxy plane (Keel 1980; Lawrence and Elvis 1982) rather than by a 
nuclear
torus. Keel et al (1994) found 80 far-IR selected Seyfert 1s and 141 
Seyfert 2s (plus some H II
galaxies that they weeded out).
So one expects a dust covering factor of $\sim$ two-thirds in this 
parameter space.
While the paper describes dust and gas covering factors as ``broadly 
consistent''
with the model, it would still be very valuable at this point to take 
advantage
of this sample and actually measure covering factor for the Type 1s based on
existing spectral energy distributions.

Figure~4 (from H\"onig \etal\ 2011) shows the quotient SED for $z>1$
individual quasars and the radio galaxy composite, which however only 
covers up
to 15 microns in the rest frame.\footnote{Buchanan et al (2006) does 
this type
of division for a $12\mu$ (better than optical, but not ideal) selected 
sample
of Seyfert galaxies. As expected, their anisotropy curve flattens at long
wavelengths, but oddly not at a value of unity.} Again this plot 
purports to give us
directly the anisotropy as a function of wavelength.

\begin{minipage}{5in}
\vskip 4truein
%\special{psfile=leipski-sed-quotient-indiv.eps hoffset=-30 voffset=-250 hscale=70 vscale=70}
\includegraphics{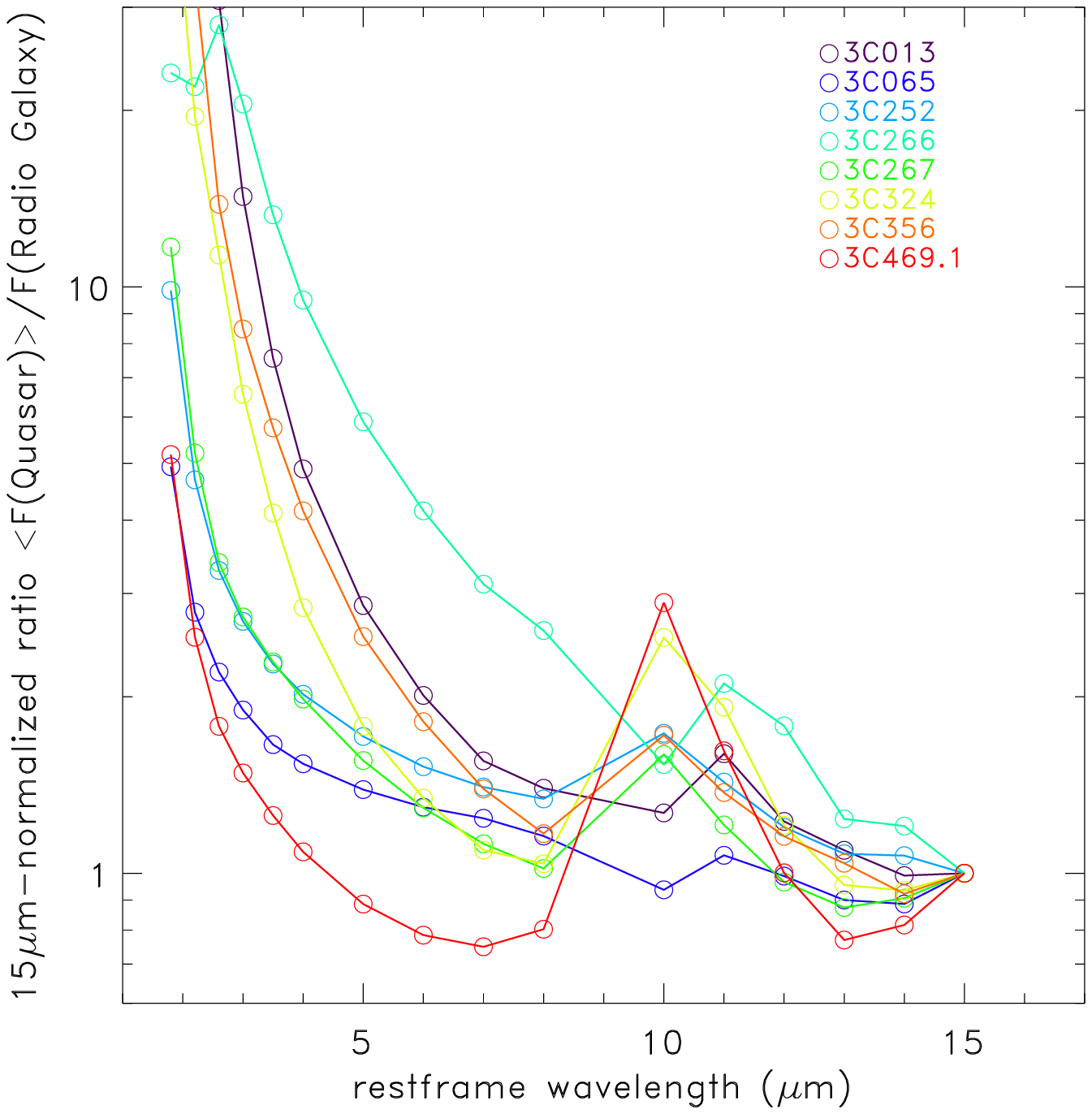}
\setlength{\baselineskip}{1ex}
\vskip 4mm
{Figure~4. The infrared SEDs of $z>1$ 3CR quasars and the matched radio 
galaxy
composite, from H{\"o}nig et al (2011).
Since these radio galaxies are all edge-on quasars, this quotient spectrum
measures the orientation dependence at each wavelength.}
\end{minipage}
\vskip 2.5mm

\section{CLASSICAL DOUBLE, OR FR II RADIO SOURCES}

\subsection{Spectropolarimetry of Radio Galaxies and the Discovery of 
Hidden Quasars}

Hidden quasars inside radio galaxies are discussed in Antonucci 1993, Wills
1999, and Tadhunter 2008. I'll describe many relevant observations here,
moving from the radio and then on up in frequency to the X-ray. First though
I will introduce some background optical information, and will return to
that waveband in Section~2.5.

It was noticed in the early 1980s that for some Narrow Line Radio 
Galaxies and
Seyfert 2 galaxies, a small measured optical polarization could often be
intrinsically large ($\gtwid10$\%) after starlight subtraction, and that the
electric vector position angle was generally perpendicular to the radio
structure axis (Antonucci 1982, 1983, 2002; Miller and Antonucci 1983; 
McLean et al 1983; Draper et al
1993 etc). Furthermore, the polarized light spectrum (similar to the 
scattered
light spectrum in spectral features) revealed the features of Type 1 
AGN, with
the first case being the powerful hidden quasar in the radio galaxy 
3CR234 shown in Fig.~2
(Antonucci 1982, 1984; Tran et al 1995; Young et al 1998). As noted earlier,
this means that the galaxies contain hidden quasars, and that their 
photons can
reach us by exiting the nuclei along the (radio) structural axes and then
scattering into the line of sight. The Tran et al 1995 polarized flux 
spectrum
is shown in Fig.~2.

Since then many
other radio galaxies, in general those with strong high-ionization lines 
like
3CR234, have shown hidden quasars in polarized light. These papers 
include most
of them: Tran et al 1995, 1998; Cohen et al 1999; Young et al 1996; Dey 
et al
1996; Cimatti et al. 1996, 1997; Ogle et al 1997 on Cygnus A (see also
Antonucci et al 1994); Cohen et al 1999; Hurt et al 1999;
Kishimoto et al 2001; Vernet et al 2001; Tadhunter et al 2002;
Solorzano-Inarea et al 2004; Tadhunter 2005.

Many Seyfert 2s were subsequently shown to have hidden Type 1 nuclei by this
method (Antonucci and Miller 1985; many references are given in Tran 
2003), but
radio loud cases were fewer, at least partially because they are more 
distant
and fainter.

While there is no question that 3CR234 hosts a powerful hidden quasar, 
it has
some properties which make it somewhat special as a radio galaxy: the 
luminous
high-ionization narrow lines, and the powerful infrared dust source (see 
Fig.~3;
also Young et al 1998). Most of the other radio galaxies
shown to host hidden quasars in this way share these properties (Cohen et al
1999). Both properties are rare in FR I radio galaxies, and at low and
moderate redshift, many FR IIs differ from this pattern as well (e.g., 
Hine and
Longair 1979; Table 2 of Cohen and Osterbrock 1981 on optical spectra; 
Ogle et al
2006, Dicken et al 2009 on infrared observations). Many of these have
low-ionization, low luminosity emission lines (Miley and Osterbrock 1979).
(All of my reference lists are undoubtedly incomplete!)

Radio galaxies with $z\gtwid1$ usually have resolved optical light, and 
much of
the light is scattered from hidden quasars (Chambers et al 1987; 
McCarthy et al
1987; Dunlop and Peacock 1993; Cimatti et al 1993). Spectacular exceptions
include 4C41.17 (Dey et al 1997) and 6C 1908+722 (Dey 1999); they show 
optical
light extended along the radio axes, but it is mostly unpolarized starlight.
However, these $z\sim4$ objects are observed at higher rest-frame 
frequencies
than explored in other objects.

\subsection{Puzzling statistics on the radio properties of FR II radio 
galaxies
and quasars}

There are a few historical papers which are interesting and illustrative 
of the
reasoning that led to much progress on unification. Peter Barthel worked
mostly on VLBI observations of superluminal sources in the 1980s, noting 
that
the beam model explained many properties such as superluminal motions 
and jet
sidedness qualitatively, but he was (according to the title of a 
rumination for
a conference) ``feeling uncomfortable'' because one had to assume that a 
large
fraction of these sources in various samples have jet axes fortuitously 
close to
the line of sight.\footnote{When selecting by high-frequency radio flux, one
preferentially selects objects whose jet axes point in our general direction
because of beaming, but this effect can't explain the apparent 
preference for
this orientation quantitatively.} Barthel later wrote a famous paper
(Barthel 1989) entitled ``Are all quasars beamed?'' suggesting that 
those quasars
whose axes lie near the sky plane
somehow fall out of quasar samples, and (inspired by the spectropolarimetry)
might be classified as radio galaxies. Note though that he did not 
entitle his
paper ``Do all radio galaxies contain hidden quasars?'' The answer to that
question would be no, but the answer to the question he posed is still 
basically
yes.

The general idea of beaming to explain superluminal motions and 
one-sided jets
was accepted by most doubters as a result of two key discovery papers 
reporting
on the so-called
lobe depolarization asymmetry. There is a very strong tendency for one 
radio lobe
in double-lobed radio quasars to be depolarized at low frequencies by 
Faraday
rotation within the observing beam on the side of the single-jet sources 
which
{\it lacks} the jet (Laing 1988; Garrington et al 1988). Most people
accepted that the depolarized lobe must be the more distant one, located 
behind
a large-scale depolarizing magnetoionic medium; thus the polarized lobe 
is on
the near side, so that the jet is also on the near side, as expected for
beaming. (A demur can be found in Pedelty et al 1989.)

The selection criteria in Laing 1988 and in Garrington et al 1988 favored
low-inclination sources; nevertheless it is amusing that the former 
paper has
this disclaimer: ``The sources observed here must then be oriented 
within about
45 degrees of the line of sight\dots to generate sufficient asymmetry in 
path
length\dots'' to fit the depolarization data. Their referees must not 
have been
particularly curious people not to ask for elaboration. We now know that 
many
of the high-inclination sources were masquerading as FR II radio galaxies.

Barthel (1989) focused on the 3CR sources in the redshift interval $0.5 
< z <
1.0$, in order to avoid low-luminosity, especially FR I radio galaxies, 
which he
did not propose to identify with quasars, noting that ``radio loud quasars
invariably\footnote{``Almost invariably'' would have been more accurate, 
e.g.,
quasar 1028+313 in Gower and Hutchings 1984} have (lobe) luminosities in 
excess of the
Fanaroff and Riley division.'' He also pointed out that the 3CR radio 
galaxies
above $z = 0.5$ have strong emission lines like quasars.\footnote{Today 
we would
say that the large majority of them have strong emission lines.} He notes
further that, based on early fragmentary data, only a subset of FR II radio
galaxies are strong IR dust emitters. Finally, Barthel crucially adds 
this claim
later in the paper: ``\dots including the $0.3 < z < 0.5$ and/or the
$1.0 < z < 1.5$ redshift range does not alter [his conclusions] markedly.''

I was particularly moved a few years later by a paper by Singal (1993),
entitled ``Evidence against the Unified Scheme for Powerful Radio 
Galaxies and
Quasars.'' I reproduce two of his figures here (the present Fig.~5 and 6).
His histograms of number densities and cumulative linear size 
distributions are
similar to Barthel's for $z > 0.5$, but he includes low-redshift FR II 
sources
in a new $z < 0.5$ bin, where ``all hell breaks loose.'' The median 
projected
linear size in quasars relative to radio galaxies no longer shows a 
satisfying
reduction expected for foreshortening, and the number density of FR II radio
galaxies becomes much higher than those of quasars.

\begin{minipage}[b]{5in}
\vskip 4truein
\begin{minipage}{2.5in}
\includegraphics{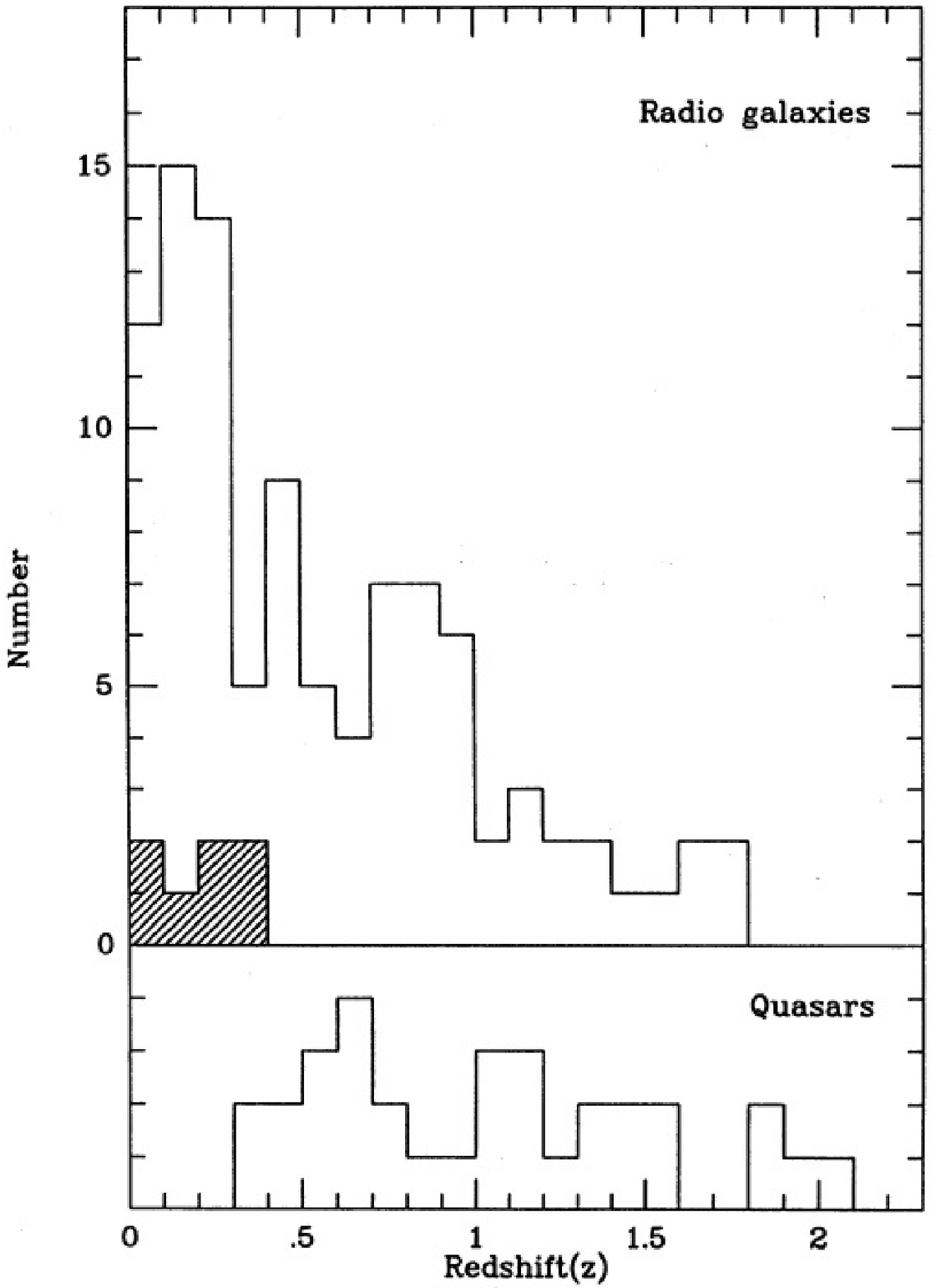}
\setlength{\baselineskip}{1ex}
{Figure 5. Redshift distribution of 131 radio galaxies and quasars in 
the 3CR
sample, taken from Singal 1993. Crucially, the hashed ``Broad Line Radio 
Galaxies''
must be mentally moved to the lower plot for the purposes of this paper. 
Note the
large ratio of the number of radio galaxies to that of quasars at low 
redshift.}
\end{minipage}
\hspace{0.25truein}
\begin{minipage}{2.5in}
\includegraphics{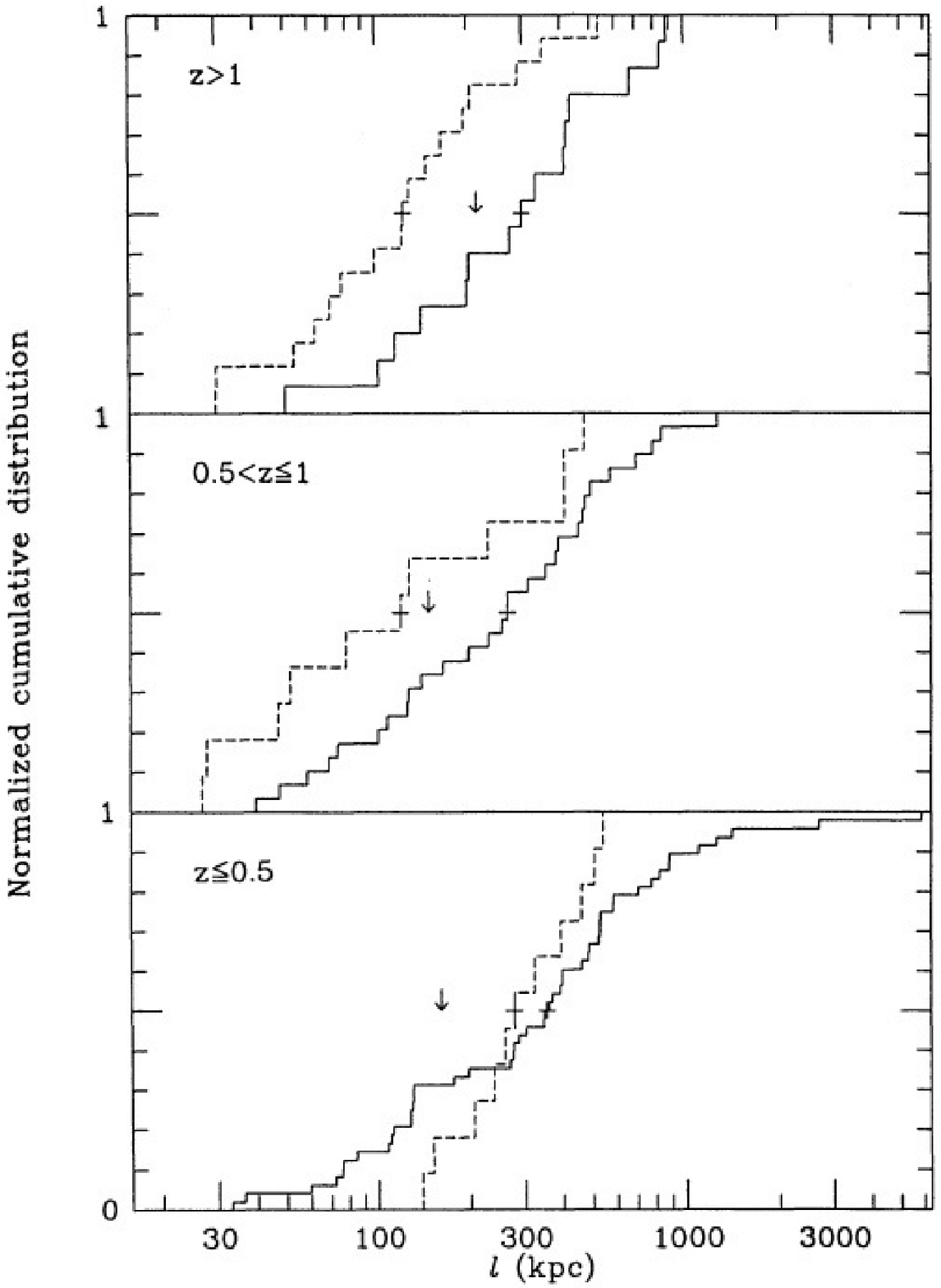}
\setlength{\baselineskip}{1ex}
\vskip 4mm
{Figure 6. Cumulative distributions of projected linear sizes of radio 
galaxies
(continuous curve) and quasars (dashed curve) in several redshift bins.
Crosses mark the median values, and arrows are for a test used by the
author, but not referred to here (Singal 1993). The foreshortening 
expected in
the shadowing Unified Model is not seen at low redshift.}
\end{minipage}
\end{minipage}
\vskip 2.5mm

Singal suggested that this spoiled the Unified Scheme, but a clever 
alternative
was suggested by Gopal-Krishna et al (1996). They showed that subject to two
assumptions justified or at least motivated by independent observations,
Singal's histograms could all be easily understood in the beam model. 
One hypothesis was
that the torus opening is set by the initial radio power of a source; 
the other was
that the luminosity of a growing giant double radio source decreases 
over time
in a certain way. Without going through all the reasoning, it turns out in
this case that in the lowest redshift bin, one is preferentially comparing
older quasars to younger radio galaxies, canceling (to the modest accuracy
attainable) the expected size difference between the two types of radio 
source.

There is another possible explanation for Singal's histograms: there is a
population of FR II radio galaxies, concentrated at low redshift, which 
simply
lacks hidden quasars.\footnote{As we will see, some FR II radio sources 
do lack
quasars, even hidden quasars, and they should form a roughly isotropic
distribution. So an easy study would be to look at the strengths of their
depolarizations. Another easy armchair ApJ Lett would be to re-create the
Singal plots, but leaving out the nonthermal galaxies (no hidden 
quasars), which
should be fairly isotropic and whose removal should make the $z < 0.5$ 
bin look
more like the other bins.}

\subsection{The infrared calorimeter for visible and hidden FR II radio 
sources}

Distinguishing these two hypotheses motivated my group and others to pursue
observations of radio galaxies in the thermal infrared, where the infrared
calorimeter (radiation reprocessed as infrared) must show us the 
putative hidden quasars, independently of
orientation. Remember that the spectropolarimetric test for hidden AGN
requires a somewhat fortuitous geometry, where a gas and dust cloud must 
have
sufficient optical depth and covering factor to reflect detectable 
light, and
it must have a view of both the hidden quasar and the observer on Earth. 
Thus
it is an incomplete method for finding hidden AGN.

David Whysong and I started imaging 3CR radio galaxies and quasars from
Singal's list at Keck Observatory in the 1990s, but equipment problems,
terrible weather, and refractory referees delayed our project until
ISO and even Spitzer were making big radio galaxy surveys, albeit at low
angular resolution. The power and elegance of the infrared calorimeter
is shown best however in the composite of high-resolution 
($\sim0.3^{\prime\prime}$) images
reproduced from Whysong's (2005) thesis, and shown here as Fig.~7.

\begin{minipage}{5in}
\vskip 3.5truein
\includegraphics{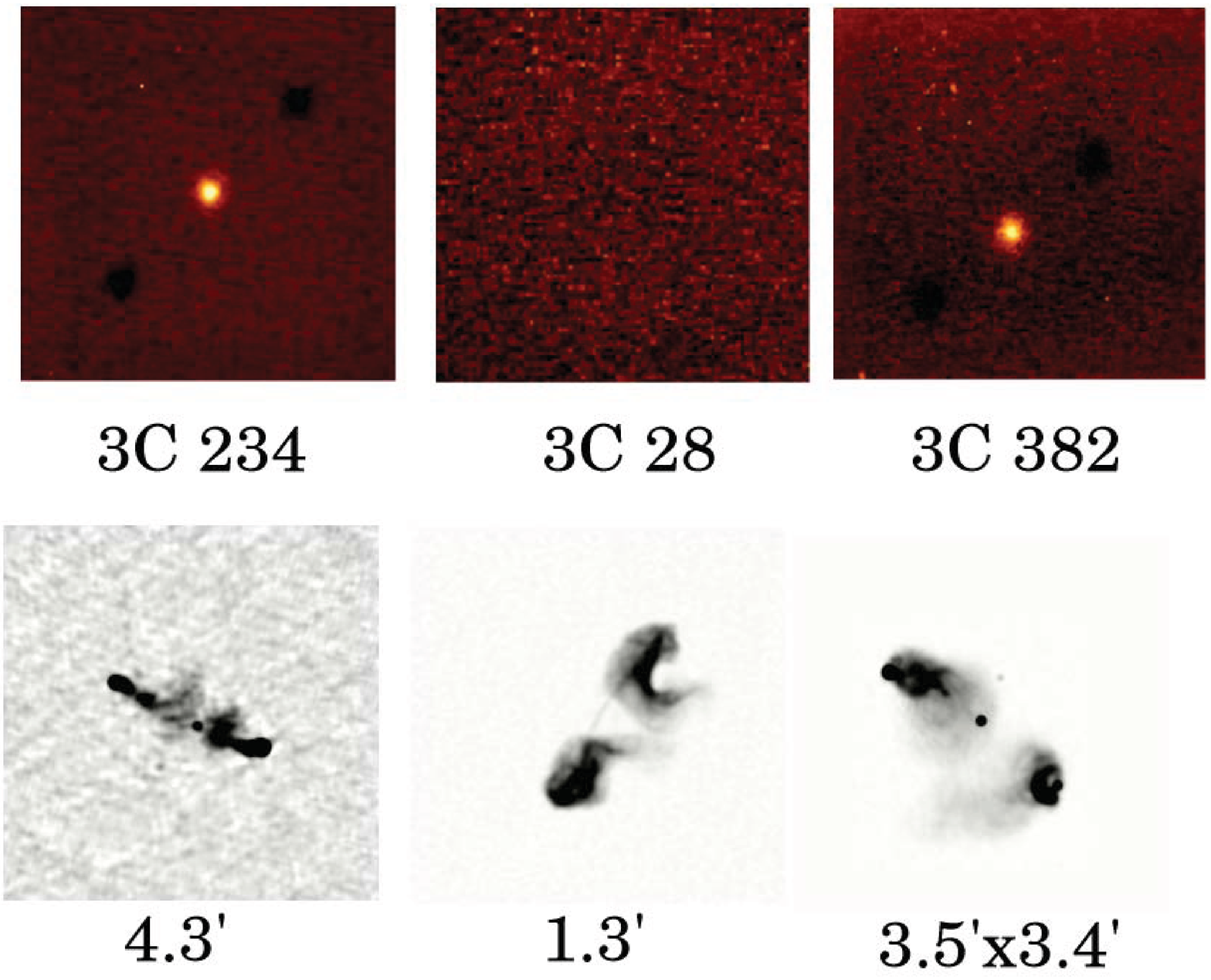}
\setlength{\baselineskip}{1ex}
\vskip 4mm
{Figure 7. Keck images of three radio-loud FR II AGN, taken in the 
thermal IR at
11.7$\mu$. 3CR382 is a visible Broad Line Radio Galaxy. 3CR234 is known 
to be
a hidden quasar from its optical polarized flux spectrum. Apparently,
3CR28 does not hide a powerful quasar since no infrared reradiation is
detected. (Note that the apparent correlation of radio core strength with
central engine type is not a general property.) From Whysong 2005.}
\end{minipage}
\vskip 2.5mm

\subsection{Infrared properties of FR II quasars and radio galaxies from 
the 3CR catalog
\label{sec:3.4}}

Many groups have been seduced by the seeming simplicity and robustness 
of the
infrared calorimeter. A rather complete review of early (pre-Spitzer)
observations is given in the Introduction to Cleary et al 2007, reporting on
Spitzer observations of 3CR objects with $0.5 < z < 1.4$. Other recent 
papers
include Meisenheimer et al 2001; Siebenmorgen et al 2004; and Whysong 
and Antonucci
2004, including some $0.3^{\prime\prime}$-resolution Keck data at 
$11.7\mu$.

Meisenheimer et al (2001)
observed 20 3CR objects with ISO, finding the results compatible with hidden
quasars, {\it except possibly for some at the low luminosity/redshift end}.
Relatedly, Siebenmorgen et al (2004) found 68 detected 3CR objects in 
the ISO archive,
finding that, ``In most 3CR objects, the mid- and far-IR flux cannot 
arise from
stars nor from the radio core because an extrapolation of either 
component to
the infrared fails by orders of magnitude.''

Shi et al (2005) used Spitzer photometry to study a sample of 3CR radio 
galaxies
and steep-spectrum quasars with good Hubble Space Telescope (HST) 
images, and
with ``a preference for $z < 0.4$''.
The data are consistent with the assertion that {\it most radio galaxies 
have hidden
quasars}. Furthermore, there is evidence for nonzero optical depths at 
$24\mu$.
Importantly, the entire infrared dust spectrum is generally inferred to 
be AGN
powered, even at $70\mu$.

Spectroscopy in the optical and the infrared can deliver more specific 
information.
Sometimes the optical narrow line spectrum indicates a lower ionization or
luminosity than is
actually present, and thus one could conclude that a hidden quasar is not
present. Again infrared photometry and spectroscopy are more complete. 
Haas et
al (2005) wrote a key paper which must be kept in mind whenever narrow line
spectra are discussed. I hope more work is done along these lines. Haas 
et al
(2005) found that in a set of seven radio galaxies and seven quasars 
from the
3CR catalog, the radio galaxies observed in the optical have on average less
luminous high ionization lines than the quasars, but that in the relatively
transparent mid-IR region, for this small sample of FR IIs at least,
{\it this is not the case}. Haas et al find that ``the luminosity ratio
[OIII]5007\AA\ / [OIV]$25.9\mu$ of {\it most} galaxies is lower by
{\it a factor of 10} than that of the quasars''!\footnote{Note that this 
large
anisotropy or at least absorption of [O III]$\lambda$5007 differs 
markedly from
the situation in the Seyfert sample discussed in Sec.~1.4.}

Similarly, di Serego Alighieri et al (1997) showed that for some 
powerful radio galaxies at least,
[O III]5007\AA\ shows up to some degree in polarized flux, indicating 
some [O
III]5007\AA\ is probably hidden by the torus in some cases; but note 
that it's
not always easy to interpret the small polarizations of [O III]5007.

Another important consideration is that at least part of the obscuration 
of the
Narrow Line Region in radio galaxies is due to the modest-optical-depth
kpc-scale dust lanes as seen so dramatically in Cyg A and Cen A. See 
discussion
in Antonucci and Barvainis 1990. Ignoring this information in testing and
exploiting the Unified Model will produce erroneous results. Absorption by
cold foreground dust is manifest as very deep mid-IR absorption features.

Now let's go back to the figures from Singal et al 1993, here Fig.~5 and 
6. We
noted that at low redshift, a {\it large fraction} of 3CR FR II radio
galaxies are {\it smaller than expected} for high inclination quasars.
This could be interpreted as evidence that there are many intrinsically 
small
radio galaxies which lack powerful hidden quasars. By powerful, I mean 
hidden
quasars roughly matched in reradiated Big Blue Bump luminosity with 
radio galaxies of the same
lobe flux and redshift. (Some allowance needs to be made for mid-IR 
anisotropy.)
Alternatively, recall that Gopal-Krishna et al (1996) cleverly noted 
that two externally
motivated assumptions would cause a universally applicable unified model 
(hidden
quasars in all radio galaxies) to lead to
data matching the observations; one has to assume that radio sources
tend to fade over time, and that the opening angle of the torus 
increases with
the {\it original} radio power. Our group was fortunate to receive time
for Spitzer IRS surveys of several types of radio-loud AGN, including 
3CR FR II
radio galaxies and quasars (Ogle et al 2006, 2007). The 2006 paper was 
the first
from a big Spitzer IRS-spectrograph survey.

These infrared data show in fact that many of the 3CR FR II radio galaxies,
especially those at $z\ltwid0.5$ (below the redshift considered in 
Barthel 1989),
have only weak and cool dust emission. Again if many of these objects have
AGN hidden by dust, then the average dust covering factor would need to 
be reasonably
large, so that the dust calorimeter should work at least statistically. 
Most 3CR
FR II radio galaxies with $0.5 < z < 1.0$ do have strong, quasar-like mid-IR
emission. This is qualitatively in accord with both Barthel's and 
Singal's figures
(which suggests a dearth of hidden AGN in FR IIs only at $z<0.5$). Also 
we now know
from Spitzer that essentially {\it all} the $z \gtwid1$ radio galaxies in
the 3CR can be unified with quasars by orientation (Ogle et al 2006; 
Haas et al 2008;
Leipski et al 2010; de Breuck et al 2010, which covers (non-3CR) 
galaxies up to
$z = 5.2$). Compare Fig.~8, showing mid infrared luminosity vs.~z, with 
Figs.~5
and 6 (from Singal 1993).\footnote{One can also see this effect to some 
extent
in the closely related top parts of Fig.~8 of Cleary et al 2007.}

\begin{minipage}{5in}
\vskip 2.75truein
\includegraphics{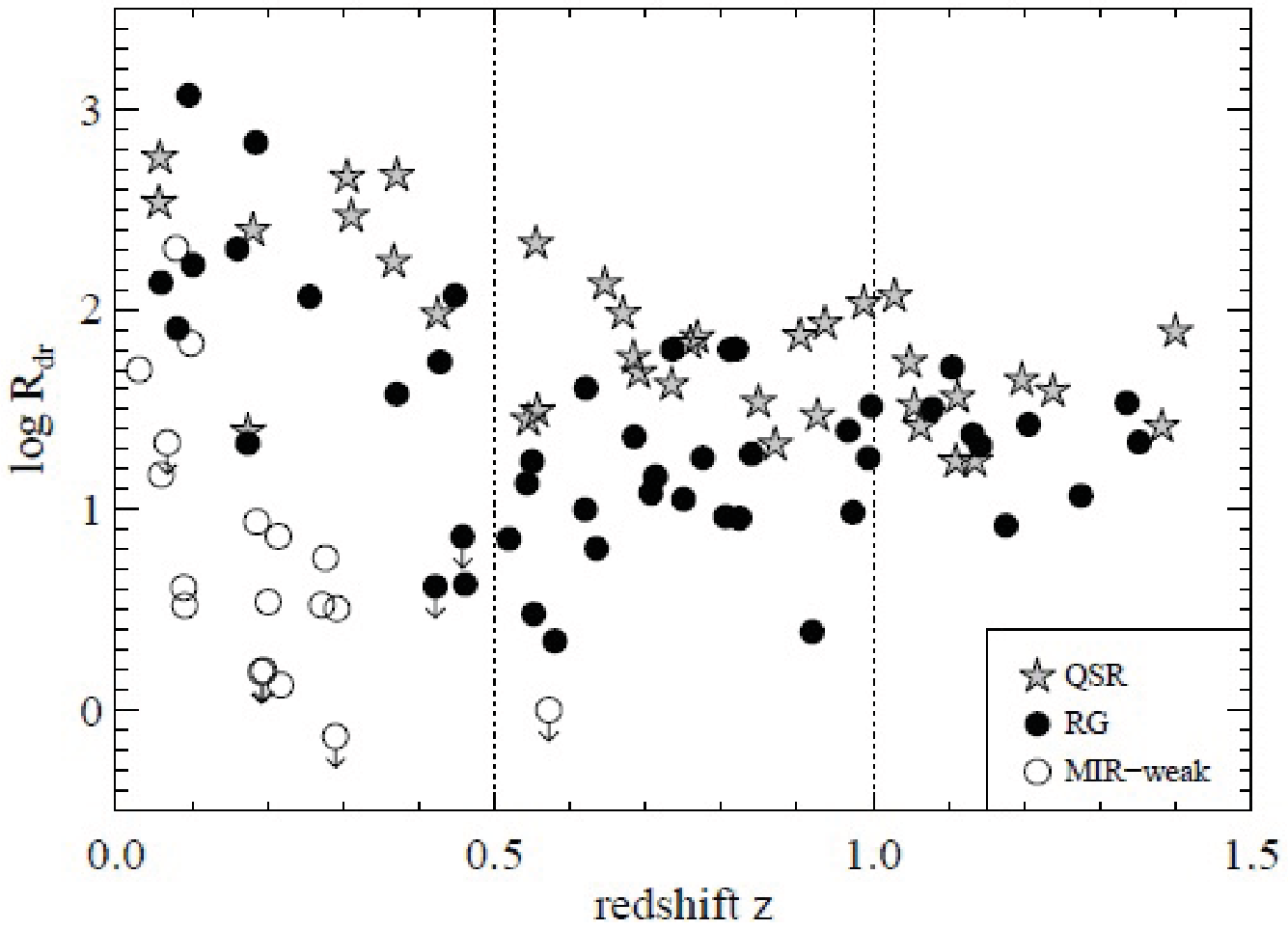}
\setlength{\baselineskip}{1ex}
\vskip 4mm
{Figure 8. The ordinate shows the rest-frame 15 shows the luminosity $\nu L_\nu$, 
normalized
to the rest-frame 178 MHz luminosity, also referring to $\nu L_\nu$. Plotting this
normalized infrared luminosity allows us to compare objects of fixed 
power in the
nearly isotropic diffuse radio emission. Thus, there are no selection 
biases with
respect to orientation and so the objects in various categories are 
directly comparable.
Stars represent quasars and filled circles refer to the putative hidden 
quasars,
with L(MIR)$\sim$L(MIR) for visible quasars. Empty circles lack visible 
or hidden
quasars, and their presence at low $z$ can explain the present 
Figs.~5--7. (see text).}
\end{minipage}
\vskip 3mm

Now let's see how the accretion luminosities, taking account of the IR data,
compare with $L_{\rm Edd}$ (Fig.~9). Among these radio galaxy black 
holes, the
ones considered to be hidden quasars (for which probably $M_{BH}\sim10^9 
M_\odot$)
radiate at $\sim0.1$\% of Eddington in dust reradiation only, and the others
are below that value. (Figure~9 is a preliminary version of this plot, 
with some
subjective elements, kindly supplied by P.~Ogle 2010.) Although this cut 
isn't always
accepted by nature, it's interestingly close to the value expected for 
the shift
from ADAF to thermal optically thick Big Blue Bump ($\dot m_{\rm 
crit}=1.3\alpha^2$ which produces
$L\sim0.13$\% of $L_{\rm Edd}$, for 10\% efficiency in $\alpha=0.1$ 
models: Esin et al 1997).

\begin{minipage}{5in}
\vskip 3.75truein
\includegraphics{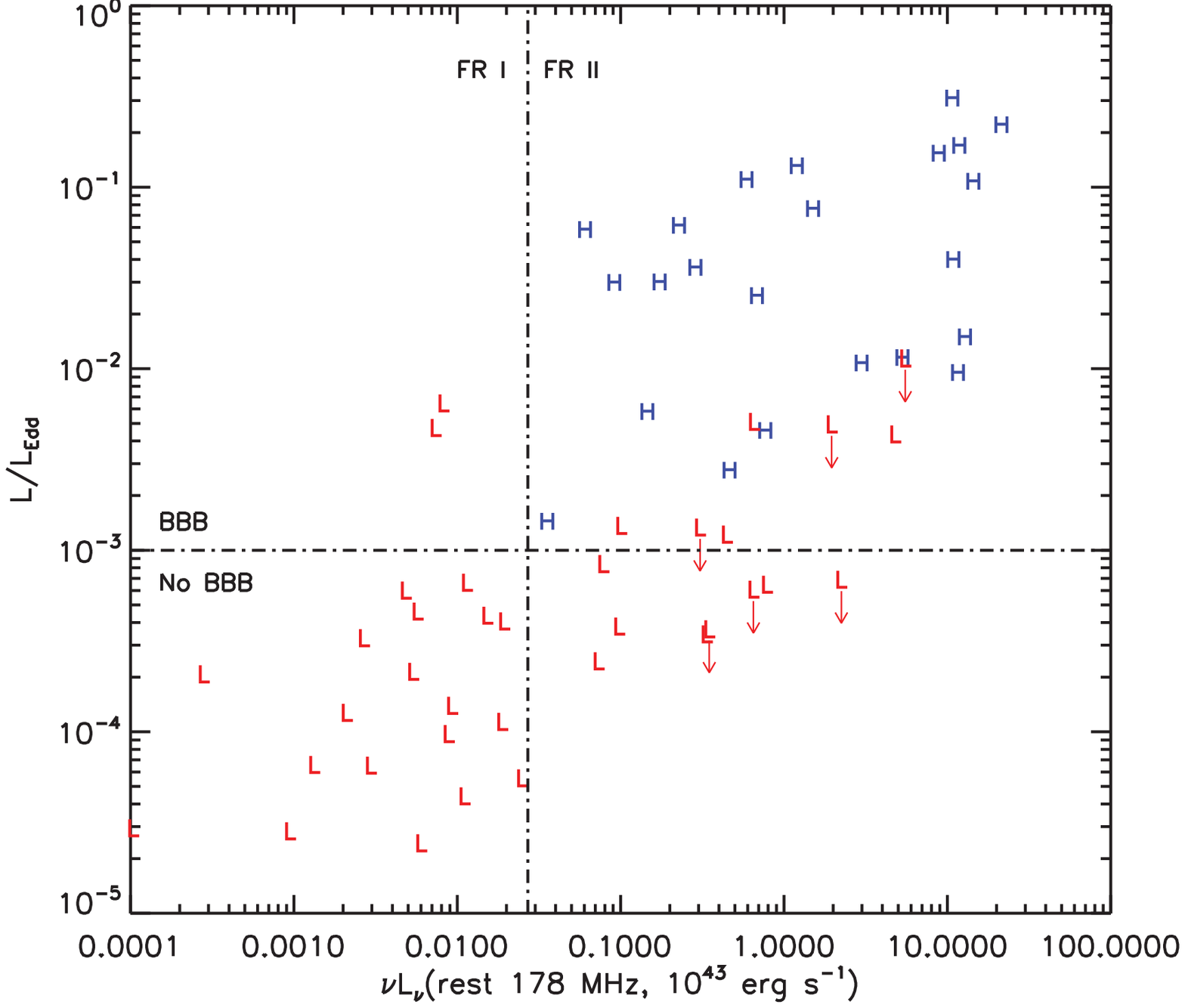}
\setlength{\baselineskip}{1ex}
{Figure 9. This plot of 3CRR radio sources is preliminary and involves some
judgment regarding covering factor, absorption inside the torus, infrared
anisotropy, and bolometric corrections. This plot, and the details about 
these
assumptions, can be found in Ogle \etal\ 2011b. Masses are based on 
spiral bulge or
elliptical host luminosity. Uncertainties are a factor of several. Note
that the objects with quasar-like mid-infrared luminosities are almost 
all of
type FR II, as well as high in ionization level. Details in Ogle et al 
2010b}
\end{minipage}
\vskip 4mm

If one were to display the apparent {\it optical} luminosities in
Eddington terms, there would be no change in narrow line ionization (see $H$
and $L$ symbols) at this expected location in $L/L_{\rm Edd}$ of the 
accretion mode change.
But using the infrared luminosities\footnote{Marchesini et al (2004) present
a proxy for this test which could be made without IR data. Using line 
emission
to estimate bolometric luminosities for radio galaxies, they found a 
probable
gap in $L/L_{\rm Edd}$ at $\sim0.01$.} seems to confirm theory 
spectacularly,
if approximately. It is also very clear that both the ionization level and
(more surprisingly) the FR type are tightly correlated
with accretion mode, as deduced by many authors over the years, and cited in
context in this paper.

When a galaxy lacks an observable Big Blue Bump, even when using the IR 
reradiation to
find it, we can of course only put an upper limit on the flux of that 
component.
According to theory, objects with $L/L_{\rm Edd}\ltwid0.01$ aren't 
expected to
produce optically thick accretion disks (Big Blue Bumps), so perhaps few 
weak
ones are missed (Begelman, Blandford and Rees 1984). However, the theory
isn't yet robust enough to be used in this way with certainty.

There are several fiducial luminosity comparisons that we can make to 
our infrared upper
limits in the radio galaxies: we can compare them to those of matched 
broad line objects
to constrain the universality of the unified model. More important 
physically,
we can compare it to the jet power. The latter is hard to get, and other 
than
for the nearby and powerful object M87, the infrared limits are too high 
to be of interest in comparing to jet
power. But for M87, any hidden quasar must produce much less radiative than
kinetic luminosity (Whysong and Antonucci 2004; Owen et al 2000; Perlman 
et al
2001).

I end this section with some general caveats on unified models and accretion
modes for FR II radio sources. 1) We're discussing the
highest-luminosity radio sources at each redshift, and in the case of the
complete geometrical unification at $z > 1$, we're only talking about 
some of
the most luminous radio sources in the universe. 2) The statistical
significance of the Singal figures makes them robust, but it isn't 
sufficient
for any further investigation by subdivision. 3) At least at the highest
redshifts, we know that there is a major contribution to the projected 
linear
sizes of radio sources besides foreshortening. It was shown by Best et 
al 2000
and subsequent papers that the $z\sim 1$ 3CR radio galaxies' projected 
linear
sizes in the extended gas correlates strongly with ionization level, the 
smaller
ones tending to have shock spectra. Aside from adding noise to the radio 
size
tests, this doesn't affect the discussion too much, though of course it's
important from a physical point of view. Note also that that correlation 
is shown
in Best et al 2000 for
the radio galaxies alone, where orientation is relatively unimportant, 
so the
claim would have to be softened if made for the entire high-$z$ 3CR sample.

\subsection{Nonthermal optical ``compact cores''}

Several authors have commented on the optical point sources seen even in 
some
Narrow Line Radio Galaxies which lack a visible Type 1 spectrum.  This 
is different
behavior than that in Seyfert 2s, few of which have point sources or 
variability ---
their mirrors are extended and spatially resolved by the Hubble Space 
Telescope
in nearby cases, e.g., Capetti et al 1995a,b; Kishimoto 1999, 2002a,b.

It's artificial to separate the FR Is from the FR IIs in this context, 
because
the entire radio galaxy population empirically separates itself in a 
different
way:  the large majority of 3CR FR Is (these are nearby, $z \ltwid0.2$), and
many low ($z < 0.5$) and some intermediate ($0.5<z<1.0$) redshift
3CR sources have optical nuclei which are consistent with emission from the
unresolved bases of the radio jets. References are given below.

In general there are no spectra available of these optical point 
sources, and
certainly no spectropolarimetry. However, the red-region HST point source
luminosities {\it and fluxes} correlate fairly well with the 5GHz (usually
flat spectrum) radio cores, which are indeed the bases of jets as shown 
by VLBI
maps. It's important that the correlation shows up in a flux-flux plot as
well as in a luminosity-luminosity plot.\footnote{In 
luminosity-luminosity plots of AGN (radio
loudness vs.\ optical power is an example of an exception), you will almost
always see a correlation because more powerful objects tend to have more of
everything. When I retire I'll make a plot of the number of bookstores 
vs.\ the
number of bars in US cities and towns --- I predict at least an 
``astronomical-quality''
correlation, which does not mean however that readers like to drink. 
Flux-flux
plots have their own peculiarities, but they are {\it different}
peculiarities than in luminosity-luminosity. Two other very common 
statistical errors that drive
me crazy are: 1. in plots of the form A vs.\ A/B (or B/A), which are 
very common, the
correlation slope, which may be intrinsically zero, will be strongly biased
towards a positive (negative) value --- it is {\it not} a discovery when 
this
happens, if the ordinate range due to errors and population dispersions 
isn't
{\it much} smaller than the range of the ``correlation.'' People actually
publish plots like this all the time, then go as far as analyzing these 
spurious
slopes and trying to extract physics from them. 2. Survival statistics 
are often
used to deal with upper limits, but most astronomical data sets violate
the key requirement for this method: that the limits have the same 
distribution
as the detections. In astronomy, we tend to have the exact opposite 
case: that
the upper limits are usually concentrated towards the bottom of the 
distribution
of detections!}

Chiaberge and collaborators have worked carefully and doggedly on these 
sources,
and FR IIs are described primarily in Chiaberge et al 2000, 2002a,b. The 
entire
data set and analysis is consistent with (but preceded!) the inferences from
the infrared. Chiaberge et al argue that the radio galaxies that fall on the
well-populated (putative) synchrotron line in the optical/core-radio plane
are likely to be nonthermal AGN. The group of radio galaxies with larger 
optical
flux than expected for synchrotron radiation then host a visible
Big Blue Bump/Broad Line Region.

These authors further suppose that any opaque tori would be larger than the
optical point sources, in which case they would block the optical light;
therefore there are no tori in most such cases. Although M87 is an FR I (or
hybrid) source, it's worth mentioning in this section that its HST point 
source
is indeed small, because it varies on timescales of 
months.\footnote{However,
this in itself is still no proof that it's smaller than any possible torus.}

Detailed study of 100 low luminosity 3CR radio galaxies with HST at
$1.6\mu$ is quite consistent with the prior conclusions of the Chiaberge 
et al
group. In particular the low ionization galaxies of both FR types show 
``central
compact cores'' which are probably nonthermal in nature (Baldi et al 2010).

Powerful supporting evidence by the same group comes in the form of a
spectroscopic survey of $z < 0.3$ galaxies (Buttiglione et al 2010).  As had
been noted by e.g., Hine and Longair (1979), FR II radio galaxies can be
naturally divided into low and high ionization objects. But Buttiglione 
et al
(2010) go a big step further and assert that their emission-line 
``excitation
index'' is bimodal (their Fig.~4)! The first and only other claim of a 
related
bimodality of which I am aware is that of Marchesini et al (2004), who 
cite such
a feature in the distribution of $L/L_{\rm Edd}$ at a value of 
$\sim0.01$. It's
worth keeping that value in mind.

Just as for the infrared data, the demographic conclusion from these optical
studies is that the fraction of 3CR FR II radio galaxies with visible or
hidden quasars is relatively small at low redshift, increasing up to at 
least
$z \sim 0.6$ (see also Varano et al 2004, and for a slightly different 
opinion,
Dicken et al 2010).

Finally, it's been found that as a group, ``radio loud'' galaxies and 
quasars
(in this case with a flux cutoff or 3.5mJy at 1.5GHz --- which is 
extremely low
compared with the 3CRs) are clustered differently, with the quasars favoring
richer environments (Donoso et al 2010). This result is consistent with 
the views
expressed here because objects with $L_\nu$ (1.5GHz) $\gtwid10^{33}$
erg/sec/Hz (essentially the range of 3CR radio galaxies and
quasars) do cluster more like quasars, according to that paper.  
Nevertheless it
strongly suggests that the shadowing unification doesn't apply at lower 
radio
luminosities.\footnote{Another easy but worthwhile ``armchair ApJ 
Letter'' could be
written to test whether restriction of the Donoso et al (2010) radio 
galaxies to those
of high radio luminosity would cause their clustering properties to 
match those
of the quasars.}

\subsection{X-rays}

Hidden radio quasars are characterized by large
($\sim10^{22}$ -- $10^{25}$ cm$^{-2}$ or more) cold absorbing columns, 
like the
Seyfert 2s and radio-quiet quasar 2s. The Low Ionization Galaxies 
generally don't show large
columns\footnote{One exception is found in Ramos Almeida, et al 2011. 
For a radio-quiet
exception, see Filippenko 1984.} and this suggests that they lack tori 
and Broad
Line Regions, although it's always possible
that the jet continuum emission extends beyond a torus (Hardcastle et al 
2009).
See their Fig.~16 and their table 7 for the
columns for Low Ionization Galaxies and other classes of AGN. These authors
argue persuasively that one can separate X-ray components from hidden 
quasars
and from jet emission, modeling the X-ray spectra with high-column and 
zero-column
components, respectively (see Figure~10).

\begin{minipage}{5in}
\vskip 2.75truein
\includegraphics{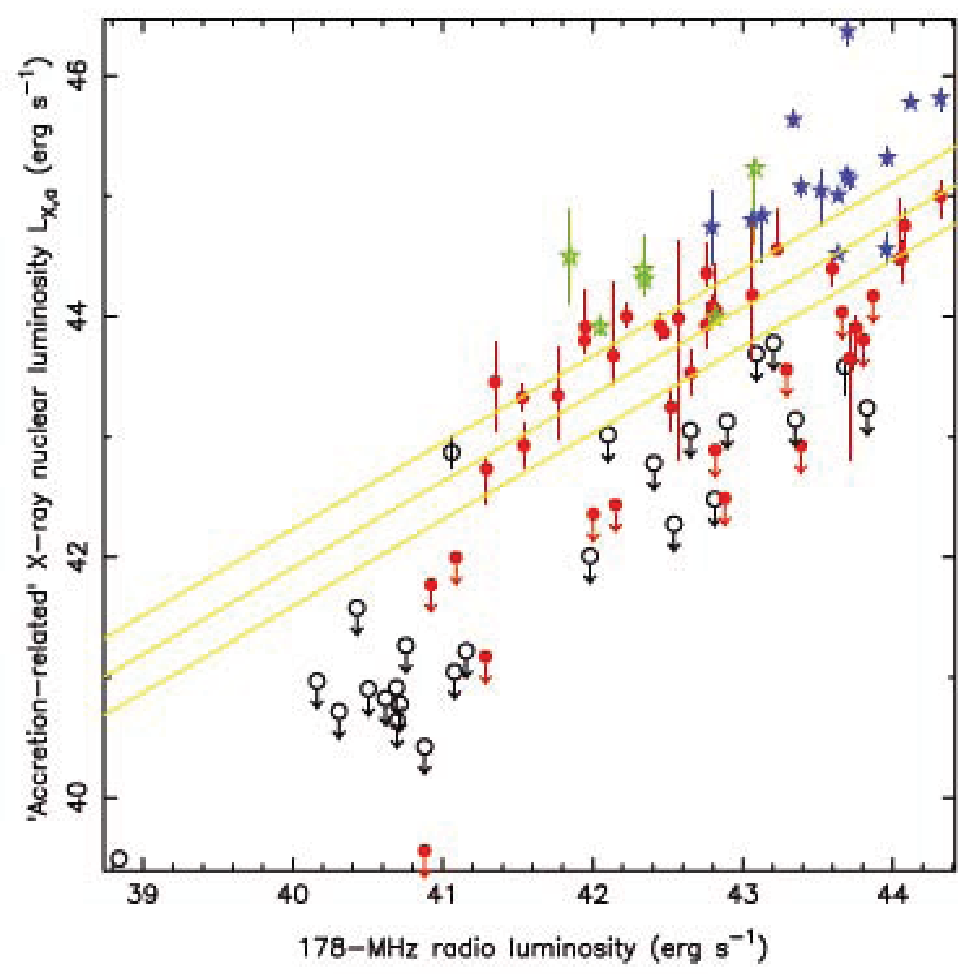}
\setlength{\baselineskip}{1ex}
\vskip 4mm
{Figure 10. The vertical axis shows the X-ray luminosity for the 
`accretion-related' (absorbed) component as
a function of 178-MHz total radio luminosity for the $z < 1.0$ 3CRR sample.
Regression is for detected Narrow Line Radio Galaxies only. Black open 
circles indicate
Low-Ionization Narrow Line Radio Galaxies, red filled circles 
High-Ionization
Narrow Line Radio Galaxies, green open stars Broad Line Radio Galaxies 
and blue
filled stars quasars. Clearly the Low Ionization objects have little or no
thermal emission. (Adapted from Hardcastle et al 2009)}
\end{minipage}
\vskip 3mm

Thus with reasonable SNR in the X-ray spectrum, one can say which spectral
component likely dominates that ostensibly derived from jet synchrotron 
emission
and those directly related to a copious accretion flow --- and often one
dominates completely. Examples of each type are shown in the FR II
galaxies in Rinn et al 2005; Kraft et al 2007; Trussoni et al 2007; and
Evans et al 2010. The X-ray spectroscopic survey
of high-$z$ 3CR objects of Wilkes et al 2009, like the infrared study
of Leipski et al 2010, is extremely supportive of complete unification of
3CR radio galaxies and quasars at $z\gtwid1$.

The putative accretion-disk K-$\alpha$ lines in the thermal radio 
galaxies and
quasars seem to be weaker and narrower than in Seyferts, a fact often 
attributed
to an inner edge of an optically thick flat accretion disk at a greater 
radius
than that at which the disks in radio quiets terminate (e.g. Ogle et al 
2005;
Sambruna et al 2009; Tazaki et al 2010). The accretion disks are often 
taken to have inner edges at
the Innermost Stable Circular Orbit\footnote{An interesting possibility 
is that
a large ISCO results from disk counter-rotation (Garofalo 2010).}, but 
this is
controversial (e.g. Agol and Krolik 2000). If the innermost part of the 
opaque
thin disk is missing, the efficiency of the standard\footnote{Optically 
thick,
geometrical thin disks with the alpha viscosity prescription, e.g., 
Pringle 1981.}
disk is reduced, and also the thermal spectrum is cooler, so one would 
expect
(but doesn't see) corresponding changes in the spectral line ratios and
luminosities (Ogle et al 2005). Also in a recent study by Molina et al 
(2008)
on a sample selected at 20--40 keV, for their six ``definite'' Broad 
Line Radio
Galaxies, ``we only find marginal evidence for weaker reprocessing
features in our objects compared to their radio quiet counterparts.''
More commentary on this can be found in Kaburaki et al 2010.

\section{FR I RADIO GALAXIES}

\subsection{Radio Properties}

These lower-luminosity ($\ltwid 2 \times 10^{32}$ erg/s Hz$^{-1}$ at 1.5GHz)
big radio doubles generally have fairly symmetric twin jets on $>1$kpc 
scale,
and the lobes are edge-darkened with no terminal hotspots.
Much of the VLBI data on FR Is (and much of it on low-luminosity FR IIs) 
come
from the group behind these references: Giovannini et al 2001, 2005;
T.~Venturi, pc, 2010. The data, though somewhat sparse on speeds especially,
are consistent with the assertion that FR I jets start out relativistic, 
with the FR Is being a
little slower than the low-luminosity FR IIs. The fraction of
sources studied in the isotropically selected 2005 sample which are
visibly two-sided on VLBI maps is $\sim30$\%, vs.\ 5--10\% in earlier 
core-flux-selected
samples. Somewhat of an update was provided by Liuzzo et al (2009), with
single-epoch data on low-frequency selected FR Is, with aggregate 
results consistent
with a single unified (beam) model.

The overall statistics on the depolarization asymmetry, another powerful
constraint on the inclination distribution, show that the effect is 
weaker than
for quasars, and probably consistent with an isotropic distribution, 
though data
are scarce. For example, Morganti et al (1997) looked at this for an FR 
I sample, and found
that the depolarization asymmetry is usually weak. Garrington et al
(1996) had come to a similar conclusion, noting however that strongly 
one-sided
jet sources have strong depolarization asymmetry. Capetti et al (1995c) 
found a
strong depolarization asymmetry in 2 out of 3 intermediate radio luminosity
(between FR I and FR II) radio galaxies. I think that more 
depolarization work
should be done, not just for the AGN field but for understanding galaxy and
cluster hot gas atmospheres, where the depolarization presumably takes 
place.

\subsection{Infrared}

Finally I'm getting to a topic that is a little bit controversial. It's not
{\it very} controversial in that everyone seems to agree that most FR Is 
have
predominantly nonthermal radio/infrared/optical and X-ray nuclei, e.g. 
M\''uller
et al 2004. But it is important not to overgeneralize, and assert a direct
connection between FR I morphology and a nonthermal engine.\footnote{It 
is easy to imagine
that the few exceptions can be attributed to the grossly different 
timescales on
which the core ($\sim1$ yr) and the lobes ($\sim10^8$ yr) are created. 
On the
other hand, the host-dependence of the luminosity cutoff, strongly suggests
environmental factors also play a role (Owen and Ledlow 1994).} I listed 
a few
exceptions in Antonucci 2002a,b, going back to the first FR I
{\it quasar}, reported in 1984 by Gower and Hutchings. (I don't know why
the more common FR I Broad Line Radio Galaxies don't impress people 
equally, but some extra cachet
seems to attach to those of quasar optical/UV luminosity.)

Recently there have been several papers modeling individual FR I nuclei as
purely nonthermal, which I think have been contradicted by subsequent 
Spitzer
spectra. For example, consider NGC6251\footnote{This is the first object 
with
good evidence of a narrow line flux variation: Antonucci 1984, Fig.~1.},
analyzed by Chiaberge et al (2003), for which the spectral index over 
most of
the infrared is inferred to be $\sim-0.6$. Leipski et al (2009) shows 
that the
mid-IR spectrum exceeds their predictions, and shows strong dust emission
features. Our conclusion from the SED is that there is hot dust with at 
least
as much flux as that due to synchrotron radiation. The aperture was 
$\sim4^{\prime\prime}$
but that may not matter much because dust emitting at a few microns must
be near a somewhat powerful optical/UV source, i.e., the nucleus. It'd 
be easy
to check from the ground. Thus the
substantial near- and mid-IR dust emission may signal a hidden thermal
optical/UV also. See however Gliozzi et al 2008, who shows that the
X-rays are likely to be dominated by the jet in NGC6251. I think a 
problem with
most published decomposition of radio galaxy infrared spectra is that they
require slopes flatter than those of Blazars, opposite to the beaming 
prediction.

A Spitzer IRS spectrum for BL Lac is shown in Fig.~2 of Leipski et al 2009:
the slope is $\sim-0.8$ between 5 and $30\mu$, considerably steeper than 
the synchrotron
slopes in most published  nonthermal models for radio galaxies. What's more,
Impey et al (1988) find that ``The spectra of Blazars steepen continuously
between $10^9$--$10^{15}$ Hz\dots the [frequency] at which the energy
distribution turns down in $\sim 2 \times 10^{11}$ Hz with a very narrow 
range
of spectral indices. Half of the Blazars with less than $10^{11}$ 
L$_\odot$ show
evidence for thermal infrared components\dots'' to which I'd add: which 
should be
much more conspicuous and thus more widespread in the high-inclination 
objects
(radio galaxies).
Moreover, ``The average Blazar spectrum is flat ($\alpha\sim 0$) at 
$10^9$ Hz
and steepens continuously to  $\alpha\sim -1.5$ at $10^{15}$ Hz. Table 4 
shows
that the infrared slopes from $3\times 10^{12}$ Hz ($100\mu$) to 
$3\times10^{14}$
Hz ($1\mu$) are {\it all} steeper than 1.'' For the SEDs of FR I synchrotron
components, with lots of infrared Spitzer data, and both with and 
without dust
bumps, see Leipski et al 2009: in our models, which feature relatively low
synchrotron contributions throughout the infrared, the slopes are 
steeper than
those found by other investigators and thus more reasonable in my opinion.

There are at least two ways around this
objection to the required flatness of the synchrotron components in some 
of the
published infrared decompositions. First, the emission from 
high-inclination may
be dominated by a slow-moving component which is for some reason 
intrinsically
flatter, and not directly related to the strong beamed component (e.g. 
Chiaberge
et al 2000). Also, the Blazar samples aren't necessarily matched to the
lobe-dominated radio galaxy samples and this could conceivably make a 
difference.
They do however include many objects with FR I diffuse radio power.

Van Bemmel et al (2004) account for most of the nonstellar radiation 
from 3CR270
(=NGC4261) with a nonthermal model, but find some evidence for a weak 
thermal
component. Our Spitzer data show a big dust bump, which covers 
$3\mu$--$100\mu$,
and dominates the infrared energetically, at least as observed in the 
$4^{\prime\prime}$
Spitzer aperture. Please see Fig.~9 of Leipski et al 2009 for our spectral
decomposition, and the location of the synchrotron component. The Big 
Blue Bump is
extremely well correlated with the Broad Line Region in AGN, and I 
consider the possible
detection of broad polarized H-$\alpha$ in this object by Barth et al (1999)
well worth following up. Again skipping ahead to the X-ray, Zezas et al 
(2005)
conclude that 3CR270 is a heavily absorbed nucleus,
$N_H\sim8\times 10^{22}$~cm$^{-2}$, far higher than most FR Is (see 
Figure~10
here; also Balmaverde et al 2006). Synchrotron is thought by Zezas et al
(2005) to contribute only $\sim10$\% of the X-ray flux.

The detailed discussion of Cen~A in Whysong and Antonucci 2004 still 
represents
our views on this controversial and somewhat complicated case. We think it
contains a hidden Big Blue Bump/Broad Line Region. Optical polarization 
imaging
is relevant for this FR I radio
galaxy. Capetti et al (2007) have measured the percent polarization of 
the HST
nuclear sources at $\sim6060$\AA\ in several FR I galaxies. Restricting 
to those
with PA errors $\le10^\circ$ (the error functions have strong tails, 
unlike the
Gaussian function), the seven remaining objects are a few percent 
polarized at
random-looking angles. Cen~A has a similarly puzzling optical (R/I band)
polarization, influenced greatly by a foreground dust lane (Schreier et 
al 1996).

In the near-IR K band however, one can sometimes see through kpc-scale 
dust lanes of
modest optical depth to the nuclear occultation/reflection region
(Antonucci and Barvainis 1990; Whysong and Antonucci 2004; see also
Bailey et al 1986). Packham et al (1996) report on both the near-IR 
polarization
and the millimeter polarization (which turns out to be crucial): the
polarization of the nucleus after various corrections is given as an 
impressive
17\% (``in the near-IR''), and exactly {\it perpendicular} to the inner
radio axis. This is expected for hidden thermal AGN rather than for Blazars.
(Refined values can be found in Capetti et al 2000.) Packham et al 
remark that
the millimeter polarization, given simply as ``zero,'' is ``not\dots
consistent\dots with that of BL Lacs.'' As noted, we believe that near-IR
observations often see through the
dust lanes, enabling us to see this very high polarization exactly
perpendicular to the radio jet, as we demonstrated with the radio galaxy
2C223.1 (Antonucci and Barvainis 1990; we used a 1-channel polarimeter [!]
but our measurement was accurately confirmed with a modern instrument).

The mere fact that the PA is constant in time for each near-IR 
observation of
Cen~A is unlike BL Lacs (or compact synchrotron sources in general), as
is the perpendicular relation to the radio jet. We also think that the 
spatially
resolved azimuthal off nuclear near-IR polarization (Capetti et al 2000) is
most consistent with scattering from a normal Type 1 nucleus.

For Cen A, we mention the X-ray spectrum here (Markowitz et al
2007). The superb Suzaku spectrum shows a column above $10^{23}$ 
cm$^{-2}$ for two
separate components, and many narrow fluorescent lines, including
Fe K-$\alpha$, like a Seyfert 2.

Going back to the mid-IR data on FR Is and FR IIs generally, an imaging 
survey of nearby
objects at $12\mu$ by van der Wolk et al (2010) revealed results which are
generally understandable and consistent with other arguments: the broad line
objects, all FR IIs, were easily detected at 7 mJy sensitivity (they 
quote $10
\sigma$!), as well as most of the High-Ionization Narrow Line Objects 
(also FR
IIs). The low-excitation galaxies of both types were not detected.

Spitzer is much more sensitive than any ground-based instrument. The current
state of the mid-IR art survey of FR Is from the IRS spectrograph is 
described
in Leipski et al 2009. Here's where there is a little more controversy. We
observed 25 FR I radio galaxies, and carefully removed the star formation
contributions as well as possible using the PAH features, and also 
removed old
stellar populations using the Rayleigh-Jeans tail of the starlight, and 
using
the AGB star features at longer wavelengths. We reached the following 
conclusions
for the 15 putative pure-synchrotron sources described in Chiaberge et 
al 1999.
Of the 15 sources with ``optical compact cores'' from the Chiaberge 
group and
others (see the ``Optical'' section below), we see four with the 
infrared dominated
by contributions from the host galaxies. In another four of the galaxies 
with
optical point sources (but probably no exposed Big Blue Bump/Broad Line 
Region),
warm dust emission dominates, and is probably at least in part due to 
hidden nuclei,
contrary to the conclusions from the optical papers. In seven cases, 
synchrotron
radiation dominates the mid-IR. The comparison to the Chabarge et al core
decompositions cannot be considered definitive however because of the larger
Spitzer aperture.

\subsection{Optical}

Some information about the optical point sources was used above to provide
context for the IR fluxes, but we must note here that these cores 
(Zirbel and
Baum 1995, 2003; Verdoes Kleijn et al 2002; Chiaberge et al 1999, 2000) have
been used to argue for synchrotron optical emission and no powerful 
hidden AGN
or tori in most FR I radio galaxies; Baldi et al 2010 is closely related.
Zirbel and Baum (1998, 2003) find that the
low-luminosity radio galaxies with detected central compact optical cores
are the ones with visible (single or highly one-sided) jets, and thus 
probably
low inclinations. This is compatible with the jet idea for the optical 
cases.

There has also been a series of papers by the Chiaberge, Capetti group
(Capetti et al 2005) on emission lines from low-luminosity radio galaxies.
These papers report that the putative synchrotron continuum is sufficient
to produce the observed emission lines, if the covering factors average
about 0.3.  They estimate the ionizing continuum from that in the UV with
power laws.
%This is tricky though because:  1) The only clouds that can
%contribute much to the covering factor are those in the jet direction due
%to beaming, and 2) the synchrotron flux as seen from Earth is less than
%that seen by the clouds for the same reason.

Another ambiguity is that the UV continuum can be very steep, and the
reddening corrections may be very large and uncertain (Chiaberge et al
2002a). Also the observed ``UV excess" in at least a few low luminosity
radio galaxies is due to starlight, according to Wills et al 2004. In that
case too the ionizing radiation can't be quantified easily.
%There has also been a series of papers by the Chiaberge, Capetti group (Capetti et al
%2005) on emission lines from low-luminosity radio galaxies, which is used to
%support the pure synchrotron model, but which can perhaps be interpreted differently.
%For example, the putative beamed continuum is said to provide sufficient
%ionizing photons for making the recombination lines, but subject to the
%assumption that the narrow line covering factor is as high as 0.3.
%This is {\it in the rest frame of the putative ionization source},
%the relativistic jet. The solid angle is reduced by a factor of order the inverse square of the
%relativistic bulk gamma factor.\footnote{The claimed {\it linear} flux and
%luminosity correlation between continuum and emission lines is not entirely
%expected since the former may be affected by beaming, while the latter is not.}
%The cool radiating gas
%needs to lie very close to the jet direction, which is somewhat unexpected,
%though the gas mass requirement is said to be modest. I'm not aware of any Cloudy-type
%calculation showing that a jet spectrum would produce the suite of detected lines.

Capetti et al (2005) also made a factor of 5 correction downward to estimate
the H-$\alpha$ emission line flux from the measured flux of the blend 
with the
[N II] doublet. The factor of 5 seems too high to me, and comes not from 
typical
AGN behavior, but from a UGC sample of ordinary LINERs. Also, there was
apparently no starlight subtraction before this factor was determined from
spectroscopy (Noel-Storr et al 2003). (The effect of this can be 
estimated from
formulae in Keel 1983.) Finally, they detected in {\it most} cases probable
broad bases to the H-$\alpha$ line, but not the forbidden lines. These 
lines are
said to be ``compatible with the broad lines seen in LINERs'' by Ho et 
al (1997).
Any Big Blue Bump accompanying these lines would probably be 
undetectable, at least with
present data, so this is consistent with (but not proof of) the presence 
of a Big Blue Bump, albeit of low
luminosity. Ho (1999,2009) has argued however that for his LINERs with 
weak broad
H-$\alpha$, the central engines are radiatively inefficient. My overall
conclusion regarding the central compact optical cores is that they are 
indeed mostly
synchrotron sources, but I don't share the same degree of confidence as the
various authors.

Let us now return briefly to the question of broad emission lines in FR 
I galaxies,
concentrating on AGN that can be observed with good contrast relative to 
the host
galaxies. The most familiar object of this type is 3C120, with a fast 
superluminal
VLBI source. Several others (Antonucci 2002a gives a brief compilation) 
are also
highly core dominated, including BL Lac itself, in which the broad lines 
need to
compete with the beamed radiation in order to be detected. This suggests 
that a
Broad Line Region is sometimes visible in an FR I radio galaxy, when 
seen at low inclination.
Falcke et al (1995) wrote a clever paper on this ``missing FR I quasar 
population.''

\subsection{X-rays}

This section will be brief because the results of many excellent studies are
simple and consistent, within the noise and the limited number of sources
analyzed, and the selection biases specific to each. Refer again to the 
present
Fig.~10 (Hardcastle et al 2009) for strong evidence of very weak 
(ostensibly)
accretion-related power Low-Ionization Galaxies, including both FR types.
Other papers are generally very supportive of (and in some ways anticipated)
Hardcastle et al 2009.

Some recent surveys with lots of FR I results: Donato et al 2004; Balmaverde
et al 2006; Rinn et al 2005; Evans et al 2006; and Hardcastle et al 2006.
Overall, the great majority of X-ray spectra of FR I radio galaxies suggest
nonthermal emission. This finds strong independent support in that most 
of these
objects (the current Fig.~10) differ from Cen A, and do {\it not} show
the high absorption columns typical of hidden AGN.

\section{SMALL SOURCES: COMPACT STEEP SPECTRUM AND GIGAHERTZ-PEAK SPECTRUM}

The radio properties of these young sources are described
briefly in Section 1, the Introduction. The classic complete review is
O'Dea 1998, while a shorter but recent review is Fanti 2009.

Recall that few can grow into large bright sources, because they are
nearly as common as the big ones (as selected in the centimeter region),
but have very short kinematic and synchrotron-aging lifetimes. Recall
also that many of the tiny kinematic ages probably measure just the
age of the current stage of activity, which may repeat many times. It's also
possible that their birthrate is extremely high, but that most fade out 
before
they grow.

The experts generally seem to agree that the radio-galaxy/quasar unification
holds fairly generally, for the well-studied very radio luminous 
population. That is,
the radio galaxies are in the thermal class. At more modest 
luminosities, there
is less information, and some hints from the infrared that this
may not be the case. If so they behave similarly to the giant doubles.

\subsection{Radio Properties}

Saikia et al (2001) provide several good arguments for small ages and
unification by geometry. On the former, kinematic and synchrotron losses 
ages
are small and generally consistent. On the latter, the authors note that the
quasars are more core-dominant in the radio, and they have more
asymmetric morphologies consistent with oppositely directed twin jets. 
Several
optical and radio papers present evidence for absorption by molecules and
HI, preferentially for the galaxies and thus near the plane according to
the Unified Model, e.g., Baker et al 2002; Gupta and Saikia 2006; and
Fanti 2009.

\subsection{Infrared}

Astronomers studied this class of radio source in the infrared with IRAS
(Heckman et al 1994) and with ISO (e.g., Fanti et al 2000). These
radio emitters show
generally high (quasar-like) power in the aggregate. The ISO mission was 
able
to make many individual detections, but the Fanti et al 2000 paper did
not make a comparison of their observed galaxies with GPS/CSS quasars.

In the Spitzer era, Willett et al (2010) presented data for eight
{\it relatively radio-faint} ``compact symmetric objects,''
which heavily overlap the GPS class. Only one was a broad-line object 
(OQ 208);
one was a BL Lac Object. Their Fig.~8 shows a plot of the Si strength
vs.\ equivalent width of the $6.2\mu$ PAH feature, demonstrating that
hidden AGN (marked by moderate Si absorption and fairly weak PAHs) probably
dominate the mid-IR emission of the galaxies in all cases. The quasar 
has Si slightly in emission,
also as expected for the unified model. However, if considered 
bolometrically,
the AGN luminosities are low (except for the quasar) and PAH features 
indicate
that star formation may contribute significant luminosity. Ionization levels
are low. The authors favor Bondi accretion or black-hole spin energy for 
most
of the galaxies, not a thermal Big Blue Bump, so in our parlance they 
would fall
into the non-thermal class.\footnote{This
refers to the overall SED, and not the infrared emission specifically.}

Our larger Spitzer survey, Ogle et al 2010, contains 13 quasars and 11 radio
galaxies from the 3CR catalog. It contains objects of substantially higher
redshift (0.4--1.0) and luminosity ($10^{34}$--$10^{35}$ erg
s$^{-1}$ Hz$^{-1}$ at 1.5 GHz and 5 GHz), as compared with the
Willett et al 2010 sample.

The radio luminosities of our sample sound much higher than most FR IIs, 
where
the FR I/II cutoff is $\sim2 \times 10^{32}$ erg sec$^{-1}$ Hz$^{-1}$
at 1.5GHz, but the
CSS and especially the GPS sources are much weaker relative to the big
doubles at low frequency. Also note that this and their small sizes
indicate that they contain much less energy in particles and fields
overall.

It is not so easy to get a complete isotropic sample of GPS sources
because they don't have dominant isotropic lobe emission. Possibly
one could select by an emission line, preferably in the infrared, after
radio classification. We took the GPS sources from Stanghellini et al
1998, a complete GHz-selected sample but likely with beaming effects
favoring low inclination. This might not be too bad since we are mainly
comparing quasars and radio galaxies, with the latter still at larger
inclinations by hypothesis (shadowing Unified Model); however they may
not be at the same level on the luminosity function. For the CSS
sources, we could find a roughly isotropic sample in the 3CR, and we took
them from Fanti et al 1995.

We find that the GPS/CSS galaxies in our sample are {\it all} powerful 
thermal dust
emitters, with vLv($15\mu$) $\sim5$--$500 \times 10^{43}$ erg s$^{-1}$.
The Si features behave somewhat erratically vs.\ optical type, although
most follow the pattern of emission for quasars and absorption for
galaxies. This can be accommodated by clumpy torus models, but since it's
specific to the small objects, it might also be from the effects of 
colder foreground
off-nuclear dust. The interpretation of the galaxies as hidden quasars
is greatly strengthened by the [Ne V] and [Ne VI] lines detected in all of
the quasars and many of the galaxies. We detect no PAH or H$_2$ features.

In summary, the infrared evidence favors a (nearly?) ubiquitous 
shadowing unified
model for the most radio-luminous small sources. Remember, however, 
these are
the highest-luminosity members of the class, and it's likely that at lower
luminosities, some radio galaxies lack the hidden uasars, based on 
Willett et al 2010.

\subsection{Optical}

Bright GPS and CSS radio sources tend to have strong line emission. All 
three
3CR CSS sources studied by Labiano et al (2005), two radio galaxies
and one quasar, show [O III]$\lambda$5007/ narrow H$\beta > 10$, so they 
have
high ionization. There is evidence for shocks also in these nice HST
spectra.

There is a wealth of information in de Vries et al 1999 and Axon et al
2000 on HST imaging of CSS radio galaxies. These groups imaged in the
H$\alpha$ or [O III] $\lambda$5007 line in tens of objects. Only the broad
line objects have point continuum sources, in accord with the Unified
Model. Thus there are almost no known ``bare synchrotron''\footnote{An
exception is PKS 0116+082, from Cohen et al 1997. This
really anomalous object has high and variable polarization like a Blazar
and good limits on broad H$\alpha$/$\lambda$5007, yet very strong narrow 
emission
lines. There is actually a
possible broad H$\alpha$ line in polarized flux, which is however
unexpected in a Blazar. This object is not analogous
to the much lower luminosity HST optical point sources studied by Zirbel
and Baum, and Chiaberge et al, and discussed in the previous two
sections. One point in common though is at least a few percent optical
polarization (Capetti et al 2007).} sources as
described above for FR I and FR II
galaxies.  This is not really a demonstrated difference however since
until very recently, only the most luminous GPS/CSS sources have been
studied in detail. In fact, very recently Kunert-Bajraszewska and
Labiano (2010) have reported on line
emission on {\it fainter} small radio sources, finding that many have Low
Ionization Galaxies spectra, just like for the giant doubles.

Both de Vries et al (1999) and Axon et al (2000) also found that the line
emission lies preferentially parallel to the radio axes, and Axon et al
add that photon counting arguments require a hidden radiation source.
Those alone are powerful arguments for unification with quasars.

\subsection{X-rays}

There are several papers on X-rays from small radio sources.
Guainazzi et al (2006) reported on a small sample of GPS galaxies at 
redshifts
between 0.2 and 1. All four with adequate SNR have large column
densities, ``consistent with that measured in High-Excitation FR II
galaxies,'' strongly indicating hidden quasars. The radio luminosities
are around $10^{34-35}$ erg s$^{-1}$ Hz$^{-1}$ at 1.5GHz, near the spectral
peaks. The 2--10 keV X-ray de-absorbed luminosities are 
$10^{44}$--$10^{45}$ erg
s$^{-1}$. (The authors give their $H_0 = 70$, but no other cosmological
parameters, so this is approximate.) Since that band covers only 0.7 dex
in frequency, the X-ray luminosity alone is $\gtwid 1 \times 10^{45}$ 
erg s$^{-1}$, and
the bolometric luminosity is likely to be at least a few times higher. 
There is
one exception, which is convincingly argued to be Compton-thick and so
opaque in the X-ray. Consistent results were reported by Siemiginowska 
et al (2008).

This group expanded their survey of GPS galaxies (Tengstrand et al 2009;
neither study included quasars) and confirms and extends these results, 
again
concentrating on the most radio-luminous objects.

\section{Summary}

First a hearty congratulations to all the theorists who predicted that
accretion would become inefficient below $L\sim 0.01 L_{\rm Edd}$. 
Apparently
the initial argument to this effect is that of Rees et al 1982. Parts of the
theory were anticipated by Shapiro et al 1976 and Ichimaru 1977. More 
recently,
Esin et al 1997 quotes 0.4 $\alpha^2L_{\rm Edd}$ as the cutoff luminosity in
the $\alpha$ prescription, which they find fits well with black hole binary
state changes.

The first radio galaxy discovered and interferometrically mapped
was Cygnus A, which has a whopping apparent brightness of 8700 Jansky
at 178MHz. At low frequencies, you can actually point the VLA 90 degrees
away from Cygnus A and map it in a sidelobe! (F.~Owen, pc 2010.)
Also interesting: considering the volume enclosed by the Cyg A distance
and the very strong cosmological evolution of FR II radio sources, folklore
says only one universe in 10,000 should be so lucky as to have such a prize!

The first map of Cygnus A is very impressive for the time, and the map
is very charming (Fig.~11 here, from Jennison and Das Gupta 1953).
There is no question that it has a hidden quasar of moderate luminosity
(Antonucci et al 1994; Ogle et al 1997; infrared fluxes from NED).
See Barthel and Arnaud 1996 to find out how such a modest quasar can be so
incredibly powerful in the radio.

\begin{minipage}{5in}
\vskip 2.75truein
\includegraphics{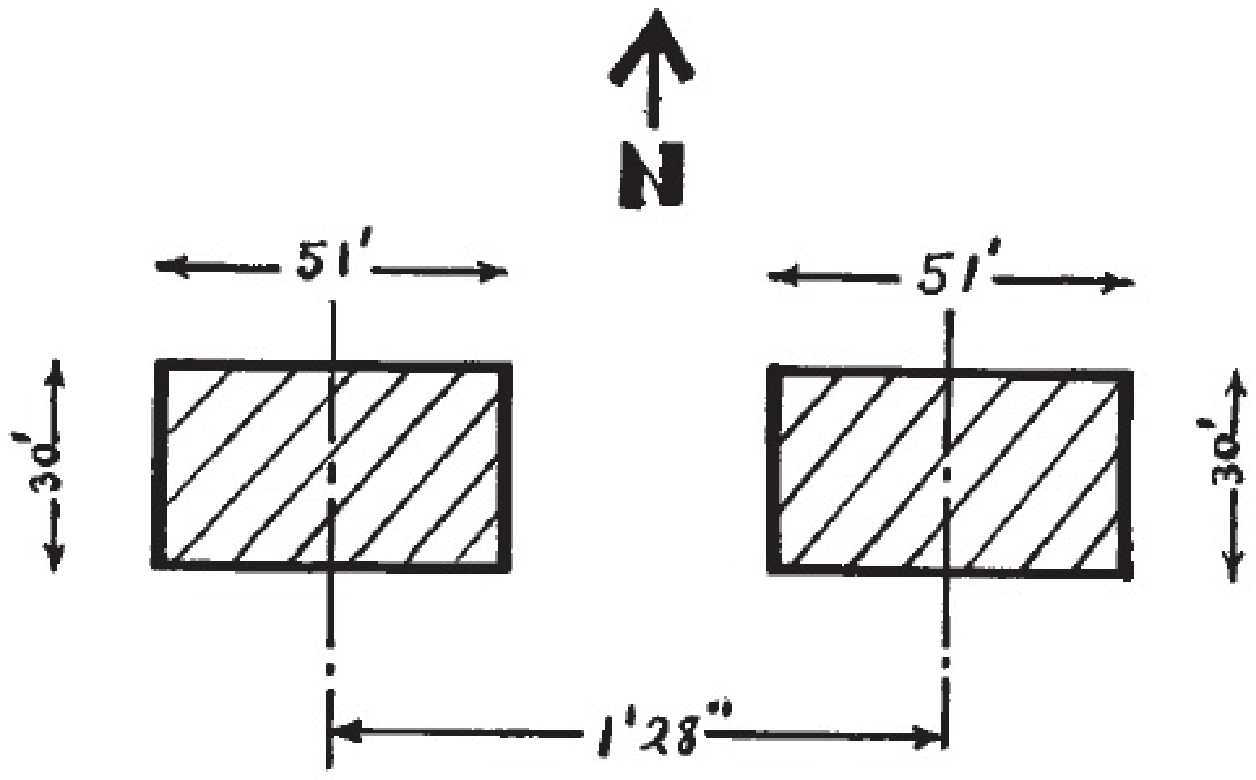}
\setlength{\baselineskip}{1ex}
{Figure 11. Approximate intensity distribution of the extra-terrestrial
radio source in Cygnus. (Jennison \& Das Gupta 1953). This very early
interferometric observation first revealed the double-lobed nature of most
radio galaxies.}
\end{minipage}
\vskip 3mm

We know now that some powerful radio galaxies, and many weaker ones,
lack detectable visible or hidden thermal AGN. It is however very important
to remember that in {\it all} but a couple of contentious cases, Type 2 
radio quiet quasars have hidden AGN,
until you get to the LINER (very low luminosity) regime, which then 
shows ADAF
behavior according to most investigators (Elitzur and Shlossman
2006; Chiaberge et al 2000; Ho 2009; many others), along with weak broad 
wings
to H$\alpha$ in many cases.

The time has come to stop proving this! We have all together quite settled
the question. It's worthwhile exploring more parameter space, e.g., weaker
radio/IR/ optical/X-ray sources, and to follow up on individual 
interesting cases, but the
overall pattern is clear now for all types of bright radio source.

So what should we do instead? The result of all this work is that we can now
hold many things constant while we vary just one thing, the tremendous
thermal emission
(Big Blue Bump). The two types must differ drastically in their structure
on relativistic scales. We are limited by our imaginations in how to take
advantage of this situation. My group is trying to determine how the
thermal/nonthermal states correlate with VLBI properties, i.e. jet
launching, collimation, and proper motion, to the extent that's possible
with current VLBI angular resolution, all while holding the large-scale 
structure
constant as far as we can discern. Another obvious observation, which
really requires next-generation X-ray telescopes to do well, is to
compare the reflection signatures of the two types. The putative 
accretion disk
K-$\alpha$ fluorescence line isn't expected to be so broad or strong if 
there
is no opaque accretion disk at small radii. Next, AGN feedback in galaxy 
evolution
might be mediated by radiation pressure on dust in many cases, or else
by PdV work or particles from jets and lobes, or other mechanisms. Here 
we can keep
everything the same (?) as far as the latter go, but turn off the radiation.
Does it make a difference? These new data might actually help with the
physics of AGN and of galaxies, and not just the astronomy of AGN.  Enjoy!

\section*{Acknowledgements}

Several astronomers have provided advice and unpublished data for this 
paper.
These include S.~Baum, M.~Begelman, P.~Best, O.~Blaes, S.~Buttliglione, 
M.~Chiaberge,
C.M.~Gaskell, M.~Gu, M.~Haas, L.~Ho, S.~Hoenig, W.~Keel, P.~Kharb, 
M.~Kishimoto,
M.~Kunert-Bajraszewska, C.~Leipski, R.~Maiolino, C.~O'Dea, F.~Owen, 
S.~Phinney,
D.J.~Saikia, A.~Singal, D.~Whysong and B.~Wills. C.~Leipski, S.~Hoenig, 
D.~Whysong and
P.~Ogle provided unpublished figures. P.~Ogle, S.~Willner, and 
R.~Barvainis suffered the most (after the author),
locating many errors and ambiguities. To the extent that the text sounds 
literate,
my editor/typist Debbie L.~Ceder deserves credit.

\newcommand{\araa}{{\it Annual Review of Astronomy and Astrophysics}}
\newcommand{\pasj}{{\it Publications of the Astronomical Society of Japan}}
\newcommand{\pasp}{{\it Publications of the Astronomical Society of the 
Pacific}}
\newcommand{\pasa}{{\it Publications Astronomical Society of Australia}}
\newcommand{\aj}{{\it The Astronomical Journal}}
\newcommand{\nar}{{\it New Astronomy Reviews}}
\newcommand{\mnras}{{\it Monthly Notices of the Royal Astronomical Society}}
\newcommand{\apj}{{\it Astrophysical Journal}}
\newcommand{\apjl}{{\it Astrophysical Journal Letters}}
\newcommand{\aap}{{\it Astronomy and Astrophysics}}
\newcommand{\apjs}{{\it Astrophysical Journal Supplement}}
\newcommand{\aaps}{{\it Astronomy and Astrophysics Supplement}}
\newcommand{\nat}{{\it Nature}}

\end{document}